\documentclass[aip, preprint
amsmath,amssymb,
reprint,
pop
]{revtex4-1}

\usepackage{graphicx}
\usepackage{dcolumn}
\usepackage{bm}

\usepackage[mathlines]{lineno}

\usepackage[utf8]{inputenc}
\usepackage[T1]{fontenc}
\usepackage{mathptmx}
\usepackage{etoolbox}

\usepackage[colorlinks,linkcolor=blue,citecolor=blue,urlcolor=blue]{hyperref}

\usepackage{mathtools}
\usepackage{amssymb}
\usepackage{siunitx}
\usepackage{setspace}
\usepackage{textalpha}
\usepackage[shortlabels]{enumitem}
\usepackage{algorithm}
\usepackage{algpseudocode}
\usepackage{derivative}
\usepackage{amsthm}
\usepackage{orcidlink}

\DeclareMathOperator*{\argmax}{arg\,max}

\DeclareMathAlphabet{\mathcal}{OMS}{cmsy}{m}{n}
\SetMathAlphabet{\mathcal}{bold}{OMS}{cmsy}{b}{n}

\def\equationautorefname~#1\null{Eq.~(#1)\null}

\makeatletter
\def\@email#1#2{%
 \endgroup
 \patchcmd{\titleblock@produce}
  {\frontmatter@RRAPformat}
  {\frontmatter@RRAPformat{\produce@RRAP{*#1\href{mailto:#2}{#2}}}\frontmatter@RRAPformat}
  {}{}
}
\makeatother

\begin{document}

\title[Local extraction of three-dimensional magnetic reconnection X-lines]{Local extraction of three-dimensional magnetic reconnection X-lines}

\author{M.\,M.\,Richter\orcidlink{0009-0009-0005-9727}}
\affiliation{Institute for Theoretical Astrophysics, Heidelberg University, 69120 Heidelberg, Germany.}
\email{max.m.richter@protonmail.com}
\author{P.\,A.\,Mu\~noz\orcidlink{0000-0002-3678-8173}}
\affiliation{Center for Astronomy and Astrophysics, Technical University of Berlin, 10623 Berlin, Germany}
\author{F.\,Spanier\orcidlink{0000-0001-6802-4744}}
\affiliation{Institute for Theoretical Astrophysics, Heidelberg University, 69120 Heidelberg, Germany.}

\date{\today}

\begin{abstract}
	Magnetic reconnection is one of the most important magnetic energy conversion processes observed in laboratory and space plasmas.
	It describes the breaking and joining of magnetic field lines, leading to the release of magnetic energy and the acceleration of charged particles.
	Finding regions where fast reconnection occurs is key to understanding this process.
	However, identifying such reconnection events within a turbulent environment in three dimensions remains a challenge.
	In this work, we develop a new framework for identifying magnetic reconnection using 3D turbulent plasma simulations.
	First, we apply bifurcation lines from fluid visualization to magnetic fields and show that they can be identified with X-lines of magnetic reconnection.
	For reconnection configurations with magnetic guide fields, we introduce a novel concept of quasi X-lines (QXL).
	Using the spatial information of X-lines in numerical simulations, we present a local technique to estimate the reconnection rate, obtaining a distribution that features a local maximum near the normalized value 0.1.
	Additionally, we provide an alternative tool to highlight current sheets in turbulent plasma by measuring magnetic shear layers as the second invariant of the shear strain tensor.
	These methods, avoiding traditional reliance on global methods, electric fields and current density, offer a new perspective to the quantitative study of magnetic reconnection in plasmas with complex magnetic field topologies.
	Validated across various plasma simulation models, including kinetic particle-in-cell (PIC) and resistive magnetohydrodynamics (MHD), our approach enables efficient exploration of magnetic field dynamics in turbulent plasma environments.
\end{abstract}

\maketitle
\section{Introduction}
\label{sec:introduction}
Magnetic reconnection is believed to be one of the most fundamental processes in astrophysical and laboratory plasmas.
It was first proposed by \citet{dungeyMotionMagneticFields1953}
and theoretically modeled by \citet{sweet14NeutralPoint1958} and \citet{parkerSweetMechanismMerging1957}.
The process describes a fast change of magnetic field connectivity, traditionally represented as a breaking and reconnecting of magnetic field lines.
Such reconnection is associated with the breakdown of ideal MHD and the frozen-in condition for the magnetic field, which may occur due to non-ideal properties of the plasma, such as non-vanishing resistivity and conversely, finite conductivity.
This process predominantly occurs in current sheets, localized regions with enhanced current density, and associated magnetic shear where magnetic energy can be stored.
Current sheets are prone to several instabilities, the most important is the tearing instability, which leads to reconnection~\citep{furthFiniteResistivityInstabilities1963}.
During reconnection, the topology of the magnetic field quickly changes, and magnetic field energy is converted to kinetic energy, heating the plasma and accelerating particles.

In various physical environments, reconnection has proven to be an important phenomenon: solar flares~\cite{janvierThreedimensionalMagneticReconnection2017} and coronal mass ejections (CME)~\cite{liThreedimensionalMagneticReconnection2021} are believed to be powered by reconnection due to a strong twisted magnetic field.
Those explosive phenomena release large amounts of electromagnetic field energy and plasma into the interplanetary space.
They constitute one of the most important drivers for transient events associated with the space weather dynamics.
Small flares (nanoflares) are also traditionally thought to be one of the possible mechanisms that could explain the solar heating problem~\cite{longcopeRelatingMagneticReconnection2015}.
Further down the interplanetary medium, reconnection takes place not only at boundaries separating plasmas of different sources, like in the Earth's magnetosphere and planetary magnetospheres in general~\cite{1968epf..conf..385D}, but also in the solar wind plasma \citep{Gosling2012}.

In particular, reconnection events take place in both the
dayside (magnetopause) and nightside (magnetotail), where they transfer plasma from the solar wind into the magnetosphere.
The release of energy by magnetotail reconnection is associated to the disturbances known as magnetospheric substorms.
They can inject energy into the ionosphere at high latitudes, finally leading to auroras in polar regions~\cite{masaakiyamadaMagneticReconnection2022}.

In laboratory plasmas, reconnection is also the cause of, e.g.,  saw-tooth oscillations in Tokamak-type fusion reactors, which are a type of periodic instabilities in plasma core temperature and density.
These oscillations can lead to instabilities in plasma confinement and subsequently impact reactor performance~\cite{chapmanControllingSawtoothOscillations2011}.

Usually all these phenomena are strongly coupled to plasma turbulence, which is ubiquitous in most of the collisionless space and astrophysical plasmas.
Reconnection can lead to turbulence but turbulence can also cause reconnection events.
The latter is mainly due to
plasma flows and interaction between magnetic structures developing in turbulence.
As a result, magnetic field lines become tangled, creating this way current sheets where magnetic reconnection can occur.
Despite many observational\citep{Phan2018,Voros2019,Wilder2022,stawarzTurbulencedrivenMagneticReconnection2022} and numerical\citep{matthaeusTurbulentMagneticReconnection1986, haggertyExploringStatisticsMagnetic2017,Franci2017, Adhikari2021, TakMagneticReconnection2021} efforts, how reconnection influences turbulence and vice versa remains a very important open question in space plasma physics \citep[see a recent review, e.g., in][]{Stawarz2024}.

The first essential step towards understanding magnetic reconnection is identifying the regions in a plasma where it actually occurs.
This is important to assess the consequences of reconnection, in particular in turbulent plasmas, where it can contribute to the dissipation and modification of the turbulent energy cascade.
Numerous methods have been established to detect this process in numerical simulations.
Most of those analyses tend to use simple visual (or partially automatized) inspection techniques based on identifying regions in a simulation with the known signatures of reconnection, such as current sheets, an X-shape magnetic field geometry with separatrices, the presence of magnetic islands (especially for elongated current sheets), the presence of plasma (electron and ion) flow with the known pattern of inflow and outflow around the X-point, a reconnection electric field perpendicular to the reconnection plane that appears as a consequence of the Faraday's law $\nabla\times\mathbf{B}=\mu_0\mathbf{J}$, (usually denoted as $E_\|$), plasma heating near the X-point region, a net dissipation of energy ($\mathbf{J}\cdot\mathbf{E} > 0$),  the inward magnetic flux transport associated with the inflow velocity, specific non-Maxwellian distribution functions, etc.\citep{Lapenta2015, Cerri2017, Califano2020, AgudeloRueda2021, TakMagneticReconnection2021, franciAnisotropicElectronHeating2022}.
Note that those approaches are purely based on physics that is independent of the dimensionality, so they can be applied in 2D and 3D.
In-situ observations of reconnection in planetary magnetospheres or solar wind also tend to use such signatures to identify reconnection in space plasmas.
In particular, by
diagnosing strong current densities correlated with parallel electric fields~\cite{stawarzTurbulencedrivenMagneticReconnection2022}, plasma outflows~\cite{vorosMMSObservationMagnetic2017} as well as plasma heating and energy transfer between field and particles~\cite{houContributionMagneticReconnection2021}.

The focus of this work is the local identification of magnetic reconnection in three-dimensional plasma simulations by using the magnetic field data only. As will be shown, our approach uses a generic algorithm that operates directly on the (typically Cartesian) simulation domain and depends only on the local magnetic vector field topology.
This enables us to identify regions of magnetic reconnection that could not be found with any of the previous methods.

In the following, we give a brief review of the physics of magnetic reconnection, together with an overview of state-of-the-art techniques for identifying reconnection in two- and three-dimensional plasma simulations.
We focus in particular on approaches that characterize reconnection through the local topology of the magnetic vector field.
Given that our method is conceptually inspired by techniques from fluid visualization, we additionally summarize the most important developments relevant for the theoretical framework of our approach.

\subsection{Local extraction of vector field features}
In the analysis of vector fields, we are often interested in the extraction of specific features.
A \textit{feature} of a vector field $\mathbf{u}$ is a subset $\Phi\subset\Omega$ of the domain $\Omega\subset\mathbb{R}^n$ of the vector field, fulfilling certain conditions $\mathcal{F}$, i.e., $\Phi=\{\mathbf{x}\in\Omega:\mathcal{F}(\mathbf{u},\mathbf{x})\}$.
Features often represent special characteristics of the vector field that are of particular interest.
These can be points, lines, surfaces, or volumes, depending on the dimension of the feature.
A common example is the extraction of the topological skeleton of the vector field, containing critical points and separatrices.

In the context of vector fields, features are classified as \textit{local} when the characteristics of all points $\mathbf{x}\in\Phi$ are determined exclusively by the field's behavior and its derivatives at $\mathbf{x}$.
Features not fitting this criterion are termed \textit{global}.
Numerical techniques have been tailored to accurately and efficiently process these local features within confined spatial domains $U(\mathbf{x})\subset\Omega$ within the overall space $\Omega$, such as the volume of a single grid cell.

A key benefit of local features is that their extraction can often be easily parallelized, as it requires only small segments of the dataset at any given moment.
For a more detailed discussion of local feature extraction in the context of visualization, we refer to \citet{hofmannLocalFeaturesTopology}, as well as \citet{rojoVectorFieldTopology2019}.
For a full mathematical discussion of vector field topology see \citet{wigginsIntroductionAppliedNonlinear2003}.

\subsection{Magnetic reconnection in two dimensions}
In 2D plasma models, a necessary condition for magnetic reconnection is the presence of so-called \textit{X-points}.
An X-point emerges when two sets of magnetic field lines approach and cross each other, forming an X-like geometry.
Mathematically, such an X-point is defined as a \textit{critical point} $\mathbf{x}_0$ where the magnetic vector field $\mathbf{B}$ vanishes, i.e., $\mathbf{B}(\mathbf{x}_0)=0$ (hence also called a \textit{magnetic null point}), and the Jacobian matrix $\nabla\mathbf{B}(\mathbf{x}_0)$ contains only real eigenvalues with alternating sign, i.e., the local topology of the field lines resemble an ``X'' due to the \textit{saddle point} behavior of the vector field.
The crossing is a topological change in the field configuration where four distinct magnetic regions meet.
These regions are divided by so-called \textit{separatrices}, special field lines that separate regions of the field with different magnetic connectivity.
In the classical Sweet--Parker or \citet{petschekMagneticFieldAnnihilation1964} models of reconnection, plasma flows in from two opposite directions towards the X-point, and springs out in two directions away from the X-point.
The magnetic flux moves with the plasma only where ideal MHD holds (the ``frozen-in'' condition).
At the X-points, even small non-ideal effects become dominant, allowing field lines to slip relative to the plasma and to reconnect.
The only other possible magnetic null point in 2D (due to the magnetic field being divergence-free) is called \textit{O-point} (center point), usually found inside magnetic islands or plasmoids caused by reconnection.
A reconnecting X-point can also contain a magnetic field perpendicular (out-of-plane) to the reconnection plane, the so-called guide field, which is very common in nature \citep{fuIdentifyingMagneticReconnection2016}.
In this scenario, usually called \textit{guide field reconnection}, the X-points are not true null points anymore, instead, only the in-plane magnetic field component form null points.

To detect magnetic reconnection in 2D, automatized methods rely on locating X-points by analyzing the eigenvalues of the Jacobian and Hessian of the vector potential and finding saddle points \citep{servidioStatisticsMagneticReconnection2010, haggertyExploringStatisticsMagnetic2017} using the previous method, but restricting the X-points to only those located on regions with strong enough values of the current density/current sheets \citep{Papini2019}, and determining not only X-points, but also the topology of magnetic islands and separatrices, e.g., by means of a contour-tree based visualization algorithm \citep{baneshTopologicalAnalysisMagnetic2020}.
In addition, semi-automatic methods have also been proposed.
For example, \citet{huIdentifyingMagneticReconnection2020} developed such an algorithm based on convolutional neural networks, in which first human experts detected magnetic reconnection events in turbulence based on the aforementioned physical signatures as a training data, so that later the algorithm could then automatically identify up to 70\% of magnetic reconnection events in turbulence successfully.

\subsection{Magnetic reconnection in three dimensions}
In contrast to the simple two-dimensional case, identifying and quantifying features linked to magnetic reconnection within fully three-dimensional plasma simulations remains a considerable challenge to this day.
The reason is that even the very definition of magnetic reconnection in 3D is far from trivial.
\citet{schindlerGeneralMagneticReconnection1988}, as well as \citet{Hesse1988}, provided the theoretical basis for the definition of 3D reconnection, which allows a much wider variety of topological reconnection configuration, such as spine-fan, torsional spine and torsional fan reconnection \citep{Pontin2011, liThreedimensionalMagneticReconnection2021}.
The most common formalism of 3D reconnection beyond global topology, as defined by \citet{priestMagneticReconnectionMHD2000}, is based on the concept of \textit{X-lines} or \textit{singular field lines}, which are the extension of 2D X-point reconnection.
They can be defined as lines that display hyperbolic behavior, i.e., a saddle-type critical point, in planes perpendicular to its line tangent.
Moreover, reconnection can only take place if there is a finite electric field $E_\|$ along the X-line. 
This is associated with a hyperbolic flow that brings magnetic flux in from two directions towards the potential singular line and carries it outwards in two other directions. We further discuss this in Section~\ref{sec:reconnection-rate}.
Despite the progress in understanding the nature of 3D reconnection, many quantitative aspects still remain unclear and a matter of debate, such as the calculation of the rate at which 3D reconnection occurs\cite{daughtonComputingReconnectionRate2014,Wyper2015}.

The simplest method to identify 3D reconnection events is using a quasi-2D approach in 3D simulations of turbulence.
This means analyzing 2D slices of turbulence on the plane perpendicular to the background magnetic field and applying previous methods, which is especially justified for strong MHD turbulence
with a strong magnetic guide field.
For example, \citet{Zhdankin2013} used a geometrical method to first determine all grid points that belong to a current sheet defined with a given threshold of the current density in 3D MHD turbulence simulations.
Note that this method can also be directly applied to identify current sheets in 2D turbulence with different plasma models \citep{Vega2020,Azizabadi2021}.
Then they identified X-points with an algorithm based on determining whether the values of the vector potential satisfy the saddle point condition in 2D.
Note that \citet{Zhdankin2013} also remarked a phenomenon already observed by \citet{priestThreedimensionalMagneticReconnection1995}:
X-points are not always located on current sheets and also vice versa, strong current sheets do not always contain X-points and thus reconnection events.
Also analyzing 3D MHD simulations, \citet{Wan2014} determined first current sheets by the method by \citet{Zhdankin2013} and then applied the method by \citet{servidioStatisticsMagneticReconnection2010} to identify X-points on those current sheets.
A different and more recent approach, applied to 3D PIC simulations, relies on finding reconnection sites via Lorentz transformations of electromagnetic fields \citep{Lapenta2021}.
Unsupervised machine learning algorithms based on clustering techniques can also automatically detect magnetic reconnection events by using several of the above-mentioned reconnection signatures \citep{Lapenta2022}.

For more general and complex 3D reconnection configurations, other techniques beyond classical vector field topology are needed.
The most popular are based on \textit{quasi-separatrix layers} (QSL) by \citet{titovQuasiSeparatrixLayersRefined1999,zhuQuasiseparatrixLayersInduced2019}.
Similar to separatrices in 2D, QSLs locally separate the flow of 3D vector fields.
However, usually QSLs are extracted numerically using global methods based on field line tracing and evaluation of the so-called Q-squashing factor, which measures flipping of field lines associated to different flow regions on the simulation boundary.
This approach is well justified in the study of coronal magnetic fields, where the natural boundary lies on the solar surface.
Unfortunately, in more general settings, such as in turbulence, the magnetic field can be very tangled and QSLs do not reach the boundary of the domain.
In such cases global methods fail.
Only recently, Wang \textit{et al.} \cite{wangMethodDeterminingLocations2023} presented a novel local method to first determine the locations of magnetic reconnection in 3D MHD simulations based on the distribution of $E_\|$ (directly proportional to the current density and thus to the magnetic field shear in the MHD plasma model) and then find X-lines, which are regions of high reconnection rates, by moving into a local reference frame and applying the general reconnection theory by~\citet{schindlerGeneralMagneticReconnection1988} and \citet{Hesse1988}.

\subsection{Reconnection rate}
\label{sec:reconnection-rate}
The reconnection rate $R_0$ quantifies the rate of change of the magnetic flux $\Phi_B$ due to magnetic reconnection, i.e., the rate at which magnetic field lines change connectivity.
The change in magnetic flux can be calculated by choosing any curve $\partial S$ enclosing a surface $S$ moving with the plasma. 
The rate of change of magnetic flux $\Phi_B$ is then given by the change of the magnetic field itself and by the movement of $\partial S$. Combined, we find
\begin{equation}
    \odv{\Phi_B}{t}=\odv{}{t}\int_S \mathbf{B}\cdot\mathrm{d}\mathbf{S}.
    \label{eq:magnetic-flux}
\end{equation}
In ideal MHD, this expression vanishes due to Alfvén's theorem, i.e., no magnetic reconnection can happen (the frozen-in condition). However, in non-ideal diffusion regions, such as present at X-lines, the situation changes.

Via Faraday's law, \autoref{eq:magnetic-flux} can be expressed as a line integral of the electric field around a closed contour bounding the surface through which $\Phi_B$ is measured, i.e.,
\begin{equation}
	\odv{\Phi_B}{t}=\oint_{\partial S} \mathbf{E} \cdot \mathrm{d}\mathbf{s},
\end{equation}
where the contour must be chosen to enclose the reconnection X-line.

In two-dimensional systems, or in quasi-2D configurations with one invariant direction (typically along the current), the contour can be taken as a rectangle with one side aligned with the X-line (reducing to a point in the strict 2D case).
The other sides contribute negligibly either because they lie in ideal-MHD regions ($\mathbf{E} \approx 0$) or because contributions cancel under periodic boundary conditions.

In fully three-dimensional scenarios, such as turbulent reconnection where X-lines may adopt complex, time-dependent geometries~\citep{schindlerGeneralMagneticReconnection1988,hesseRelationReconnectedMagnetic2005,comissoValueReconnectionRate2016}, the situation is more complicated.
It can be shown that the contour integral of the electric field, is directly related to the line integral of the parallel electric field,
\begin{equation}
	\int E_{\parallel} \,\mathrm{d}s,
	\quad \text{with}\quad
	E_{\parallel} = \frac{\mathbf{E} \cdot \mathbf{B}}{\|\mathbf{B}\|}.
\end{equation}
Here, $E_{\parallel}$ represents the non-ideal electric field, which is nonzero in the diffusion region where the frozen-in condition breaks down.
A finite value of $\int E_{\parallel} \,\mathrm{d}s$ along field lines crossing the diffusion region is the defining signature of a finite-$\mathbf{B}$ global reconnection process in the sense of \citet{schindlerGeneralMagneticReconnection1988}, typically driven by inductive electric fields.

It is important to note that $E_{\parallel} \neq 0$ can also arise from processes unrelated to reconnection, such as electrostatic fields.
However, many such contributions (e.g., from localized charges) integrate to zero over $\int E_{\parallel} \,\mathrm{d}s$, corresponding to a localized, non-global reconnection without large-scale topological changes~\citep{schindlerGeneralMagneticReconnection1988}.
Similarly, current sheets that are not actively reconnecting may exhibit finite $\int E_{\parallel} \,\mathrm{d}s$ due to other non-ideal effects, but such signatures will not be present along field lines associated with X-lines, which appear only once reconnection has started.

Following this reasoning, we can give an estimation of the reconnection rate by means of
\begin{equation}
	R_0=\frac{\mathrm{d}\Phi_B}{\mathrm{d}t}=\oint\mathbf{E}\cdot\mathrm{d}\mathbf{s}\approx\frac{1}{L}\int_{\text{X-line}} E_{\|}\mathrm{d}s,
	\label{eq:reconnection-rate}
\end{equation}
where $L$ is the length of the X-line.
Note that this is not a strict equality, but an estimation that becomes exact under the simplified scenarios mentioned above
and that can be quantified with the methods shown in this paper.
In order to compare different reconnection systems, it is further more common to compute the normalized reconnection rate $R$ as
\begin{equation}
	R=\frac{R_0}{B_{\text{in}} V_A/c},
	\label{eq:relative-reconnection-rate}
\end{equation}
where we scale the reconnection rate with the Alfvén speed $V_A$ and the inflow magnetic field strength $B_{\text{in}}$.
The Alfvén speed is given by
\begin{equation}
	V_A=\frac{B_{\text{in}}}{\sqrt{4\pi \rho_{\text{in}}}},
\end{equation}
with the inflow mass density $\rho_{\text{in}}$.
Note that those formulas are in CGS units.
The normalized value is important because most reconnection events in nature exhibit a normalized reconnection rate near 0.1,
confirmed as well by most numerical simulations with a wide variety of plasma models \citep{comissoValueReconnectionRate2016}.

\subsection{Local features in fluid visualization}
The study of vector fields and their visualization is traditionally motivated by fluid dynamics.
Here, identifying and visualizing characteristic topological and geometrical features of vector fields has been of interest for a long time.
Finding critical points and separatrices to build the topological skeleton on vector fields is one of the most established approaches to visualize fluid dynamics.
However, the flow field of a three-dimensional fluid may lack any kind of critical points while still creating highly complex behavior.

In fluid visualization, two of the most prominent structures beyond classical vector field topology are the so-called \textit{vortex core} and \textit{bifurcation lines}, which are 1D manifolds of local elliptic or hyperbolic behavior, respectively.
Numerous studies have explored the extraction of vortex core lines, applying them to fluid simulations and magnetic fields.
Prominent examples to identify vortex core lines are the vortex-criterion introduced by \citet{huntEddiesStreamsConvergence1988} or the $\lambda_2$-criterion by \citet{schafhitzelTopologyPreservingBasedVortex2008}.
\citet{schafhitzelVisualizingEvolutionInteraction2011} also studied the evolution and interaction of vortex regions and shear layers in time-dependent flow, and defined the so-called $I_2$ value to measure shear and to localize layers of high shear in velocity fields.
As will be shown, applied to magnetic fields, these shear layers indicate regions of high current density, the current sheets.

Among the most successful methods for the local extraction of line-features in vector fields is the parallel vectors operator (PV operator) by \citet{peikertParallelVectorsOperatora1999}.
Its success is mainly driven by its independence of the underlying grid geometry, so even complex unstructured grids can be used while computations remain easily parallelizable.
Originally, it was invented for the visualization of vortex core lines~\cite{sujudiIdentificationSwirlingFlow1995}, ridge and valley lines~\cite{eberlyRidgesImageAnalysis1994}, as well as separation and attachment lines~\cite{kenwrightAutomaticDetectionOpen1998}.

\citet{machadoLocalExtractionBifurcation2013} introduced an algorithm based on the parallel vectors operator for identifying bifurcation lines and saddle connectors within vector fields, including those encountered in fluid dynamics simulations and solar coronal magnetic fields, though without explicitly addressing magnetic reconnection.
Despite their potential significance, bifurcation lines have generally remained underappreciated in the scientific analysis of magnetic fields.
Notably, bifurcation lines exhibit remarkable similarities to singular field lines or X-lines.
However, to the best of our knowledge, their usage has not yet been explored in the context of magnetic reconnection in plasmas.

In the following Section~\ref{chap:methods}, we give a brief introduction to the theory of bifurcation lines (Section~\ref{sec:bifurcation-lines}), as well as their numerical extraction (Section~\ref{sec:parallel-vectors-operator} and Section~\ref{sec:extraction-of-bifurcation-lines}).
To overcome shortcomings of previous methods, we provide a new type of line feature, based on bifurcation lines, called quasi X-lines (Section~\ref{sec:qxls}).
Using these lines, we present a local method to estimate the reconnection rate in Section~\ref{sec:relative-reconnection-rate}.
We additionally show how $I_2$ shear layers can improve the localization and analysis of magnetic reconnection in current sheets (Section~\ref{sec:shear-layers}).
In Section~\ref{sec:implementation} we discuss implementation and visualization details.
In Section~\ref{chap:results}, we provide three physical plasma simulations as a benchmark for our approach and come to a conclusion in Section~\ref{chap:discussion}.

\section{Method}
\label{chap:methods}
\subsection{Bifurcation lines}
\label{sec:bifurcation-lines}
\begin{figure}
	\centering
	\includegraphics[width=0.49\linewidth]{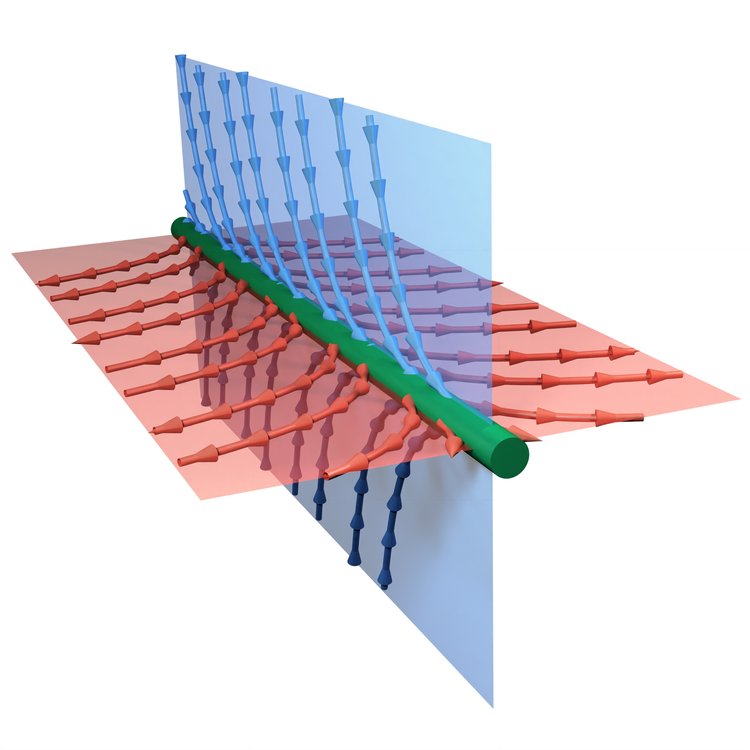}
    \includegraphics[width=0.49\linewidth]{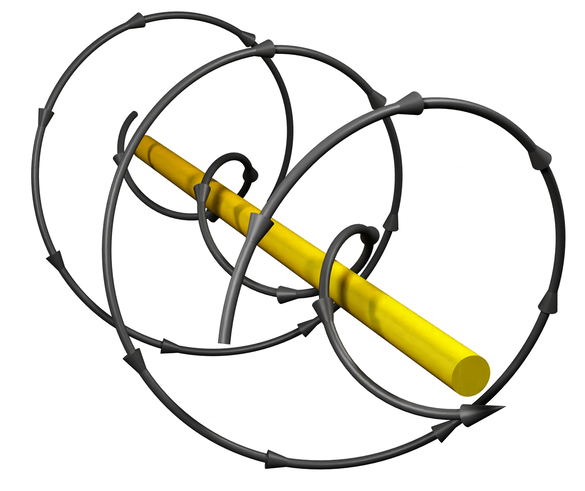}
	\caption{(Left) Example of a bifurcation line in green, which is the intersection of two manifolds of converging (blue) and diverging (red) field lines. (Right) Example of a vortex core line in yellow.
		Local rotating field is shown by two field lines in gray.}
	\label{fig:bifurcation-line}
\end{figure}
As already discussed in the Introduction, the natural generalization of X-points from 2D to 3D geometries are X-lines.
These are lines of hyperbolic, X-type behavior in the planes perpendicular to the line tangent, called the \textit{in-plane} vector field.
In fluid visualization, such lines are known as bifurcation lines, introduced by Perry and Chong~\cite{perryDescriptionEddyingMotions1987} to describe diverging field line behavior in the flow of three-dimensional fluids, similar to vortex core lines for rotating field line behavior (which are the extensions of O-points, see Section~\ref{sec:vortex-core-lines}).
As we will show, applied to the magnetic vector fields of three-dimensional plasma, these bifurcation lines can be identified with the X-lines of magnetic reconnection.

Similar to vortex core lines, a bifurcation line is a field line of the vector field and can be defined by the \citet{sujudiIdentificationSwirlingFlow1995} criterion.
Each point of the line has to fulfill the following conditions:
\begin{enumerate}[(i)]
	\item $\mathbf{B}$ is parallel to an eigenvector $\mathbf{v}_i$ of the gradient $\nabla \mathbf{B}$ (Sujudi and Haimes)
	\item All eigenvalues $\lambda_i$ of the Jacobian matrix $\nabla \mathbf{B}$ are real
	\item The major and minor eigenvalues $\lambda_1$ and $\lambda_3$ are of opposite sign.
\end{enumerate}
As an consequence of conditions (ii) and (iii), a bifurcation line is the cross-section of two local manifolds of converging and diverging field lines of $\mathbf{B}$.
An example of such a bifurcation line is illustrated in \autoref{fig:bifurcation-line} (left). The term ``bifurcation'' line is motivated by these two planes of streamlines, but is unrelated to the bifurcation theory of critical points in dynamical systems.

\subsection{Vortex core lines}
\label{sec:vortex-core-lines}
The three-dimensional extension of O-points is the set of curves known as \textit{vortex core lines}.
In fluid dynamics, these lines represent the axes of rotation within a vortical flow.
A widely used formulation, due to Sujudi and Haimes, defines vortex core lines as field lines satisfying the following criteria:
\begin{enumerate}[(i)]
    \item $\mathbf{B}$ is parallel to an eigenvector $\mathbf{v}_i$ of the gradient $\nabla \mathbf{B}$ (Sujudi and Haimes)
    \item Two eigenvalues $\lambda_i$ of the Jacobian $\nabla \mathbf{B}$ form a complex-conjugate pair.
\end{enumerate}
When applied to magnetic fields, vortex core lines identify the central axes of flux ropes and plasmoids in elongated current sheets. An example of a vortex core line can be seen in \autoref{fig:bifurcation-line} (right).

Due to their close structural similarity, both vortex core lines and bifurcation lines can be extracted using the same computational framework.
This enables the simultaneous identification of vortex core lines as a valuable byproduct of the bifurcation line extraction of reconnection X-lines, further discussed in the following.

\subsection{The parallel vectors operator}
\label{sec:parallel-vectors-operator}
\begin{figure}
	\centering
	\includegraphics[width=0.99\linewidth]{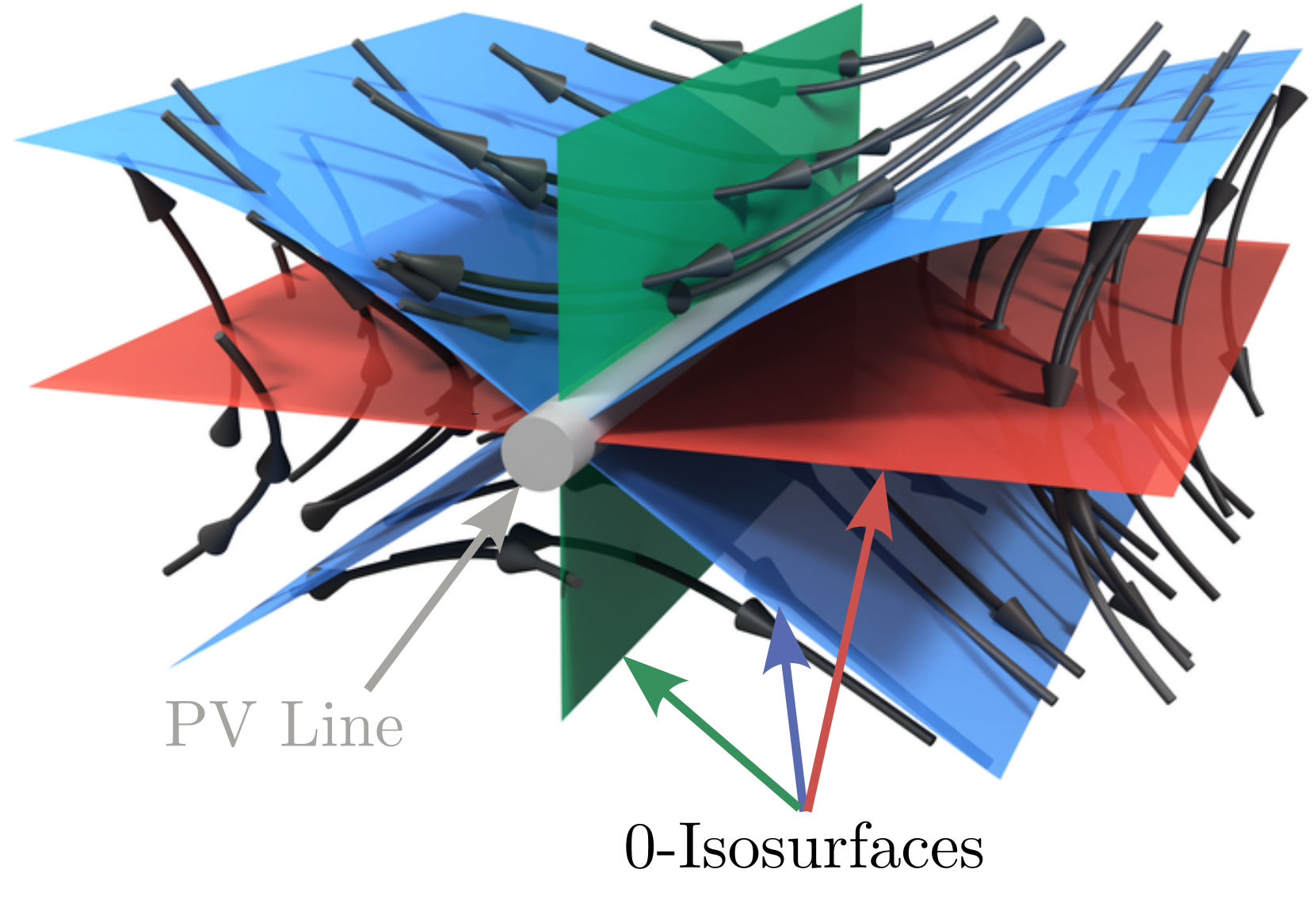}
	\caption{Illustration of a parallel vectors line (gray) as the cross-section between the three 0-isosurfaces (red, green and blue) of the vector cross product.
		Vector field lines are depicted as black arrows.}
	\label{fig:isosurfaces-ospray-big}
\end{figure}

\citet{peikertParallelVectorsOperatora1999}, and later \citet{machadoLocalExtractionBifurcation2013}, showed that vortex core lines and bifurcation lines can be extracted simultaneously using the parallel vector operator.
Given two vector fields, the \textit{parallel vectors operator} returns the locations where the two vector fields are parallel.
Let $\mathbf{v,w}$ be two vector fields, the parallel vectors operator is then denoted by $\mathbf{v}\| \mathbf{w}$ (we say ``$\mathbf{v}$ parallel $\mathbf{w}$'') and returns the set
\begin{equation}
	S = \{\mathbf{x}\in\Omega:\mathbf{v(x)}=0\}\cup\{\mathbf{x}\in\Omega:\exists\lambda,\mathbf{w(x)}=\lambda \mathbf{v(x)} \}.
\end{equation}
In three dimensions, this set can be calculated as the vector cross product between the two vector fields, i.e., the set
\begin{equation}
	S=\left\{\mathbf{x}\in\Omega:\mathbf{v(x)\times w(x)}=0\right\}.
	\label{eq:pv-criterion}
\end{equation}
The solutions to this operator in three dimensions are closed lines, which can be understood as the intersection of the three isosurfaces defined by the equations of the vector cross product (See \autoref{fig:isosurfaces-ospray-big}).
We call these lines the raw \textit{parallel vectors lines} (PV lines).

The parallel vectors operator can be used to evaluate the Sujudi--Haimes criterion, i.e., condition (i) in Section~\ref{sec:bifurcation-lines}.
However, instead of calculating the eigenvectors of the vector field explicitly, we can evaluate the parallel vectors operator with the vector field $\mathbf{v=B}$ and its convective acceleration $\mathbf{w}=\mathbf{(B\cdot \nabla)B}$.
The convective acceleration usually describes the time-independent acceleration of the vector field (with respect to space) that a particle in a flow would experience.
It is derived from the Navier--Stokes equation of fluid dynamics, however, this quantity is a feature only of the vector field itself and is hence \emph{not} dependent on the fluid interpretation of the operator.
In case of magnetic fields, this quantity is better known as the \textit{magnetic tension} and measures the local curvature of the vector field.
Its expression can be equivalently described by
\begin{equation}
	(\mathbf{B} \cdot \nabla) \mathbf{B}=\sum_j\frac{\partial B_i}{\partial x_j}B_j=(\nabla \mathbf{B})\mathbf{B}.
\end{equation}
Suppose the vector field $\mathbf{B}$ is an eigenvector of its own Jacobian $\nabla\mathbf{B}$, then it must hold that
\begin{equation}
	(\nabla\mathbf{B})\mathbf{B}=\lambda\mathbf{B},
\end{equation}
so if $\mathbf{B}$ is parallel to an eigenvector of $\nabla\mathbf{B}$, it is manifestly also parallel to its tension/acceleration.
This means that, instead exhaustively computing the eigenvectors of the Jacobian at every point using $\mathcal{O}(n^3)$ steps for the Sujudi--Haimes criterion (i), we can equivalently just compute the Jacobian matrix and multiply it with the vector field, requiring only $\mathcal{O}(n^2)$ steps.
The resulting lines will be the same.

Following this reasoning, the locations of such lines admit a direct physical interpretation: they correspond to connected loci where the magnetic field vector $\mathbf{B}$ is everywhere parallel to its own magnetic tension $(\mathbf{B} \cdot \nabla) \mathbf{B}$,
\begin{equation}
	\mathbf{B} \,\|\, (\mathbf{B} \cdot \nabla) \mathbf{B},
\end{equation}
or where either quantity vanishes.
Crucially, this definition is not restricted to numerical evaluation. For sufficiently simple magnetic field configurations, the parallel-vectors operator can be evaluated analytically, providing exact closed-form expressions for these lines. This analytical accessibility strengthens the physical interpretation and allows direct comparison with numerical results in more complex systems.

\begin{figure*}
	\centering
	\includegraphics[width=0.99\linewidth]{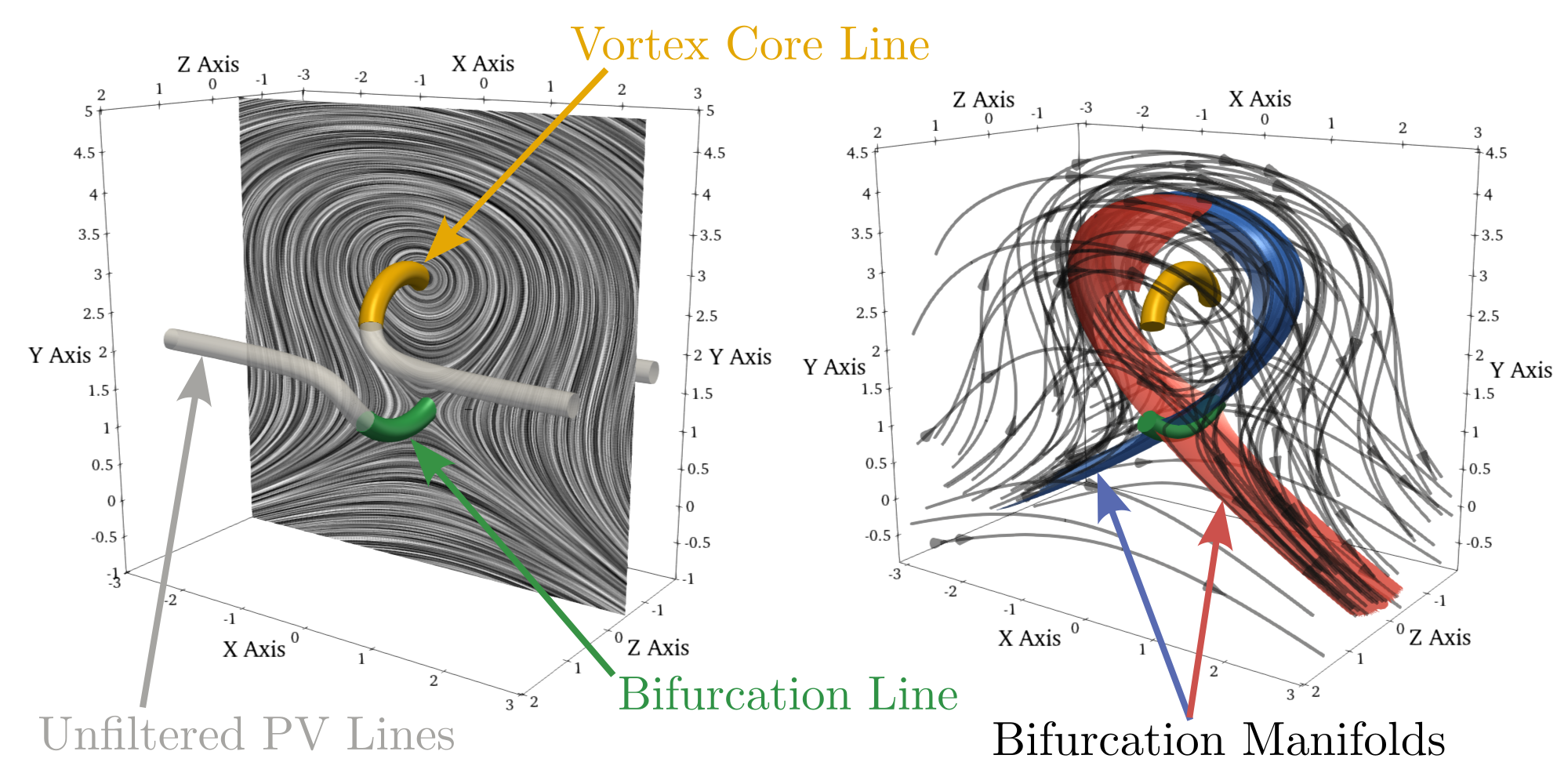}
	\caption{(Left) Unfiltered solution of the parallel vectors operator (gray) for solar flux rope model with $B_0=0.1$, where the bottom solution belongs to the bifurcation line (green) and the top line belongs to the vortex core line (yellow).
	The in-plane magnetic field is visualized using Line Integral Convolution (LIC)\cite{cabralLineIntegralConvolution2023} in a $z=\text{const}$ plane.
	(Right) Bifurcation manifolds of attracting (blue) and repelling (red) field lines.
	Some magnetic field lines are shown as black field lines.}
	\label{fig:singular-line-bifurcation}
\end{figure*}

As a concrete example, we consider the analytical model for X-line reconnection in a three-dimensional twisted solar flux rope presented by \citet{priestMagneticReconnectionMHD2000}. Such flux ropes are believed to occur in solar flares and to be driven by magnetic reconnection processes, even in the absence of magnetic null points (see Introduction~\ref{sec:introduction}). The magnetic field in this model, at a point $\mathbf{x}=(x,y,z)^\top\in\Omega\subset\mathbb{R}^3$, is given by
\begin{equation}
	\mathbf{B}(\mathbf{x})=\begin{pmatrix}
		(y-2)^2 - \mu + z^2 \\
		-x                  \\
		B_0
	\end{pmatrix},
	\label{eq:solar-flux-rope}
\end{equation}
where $\mu$ is a free parameter and $B_0$ denotes the uniform guide field. In planes $z=z_0$, this vector field has center points at $(0,2+(\mu-z_0^2)^{-1})$ and saddle points at $(0,2-(\mu-z_0^2)^{-1})$, given that $z_0^2<\mu$. The Jacobian of this system and the magnetic tension are given by
\begin{equation}
	\nabla \mathbf{B}=\left(\begin{array}{ccc}
			0  & 2 y-4 & 2 z \\
			-1 & 0     & 0   \\
			0  & 0     & 0
		\end{array}\right)
\end{equation}
and
\begin{equation}
	(\mathbf{B} \cdot \nabla) \mathbf{B}=\left(\begin{array}{c}
			2B_0 z-x(2 y-4) \\
			\mu-z^2-(y-2)^2 \\
			0
		\end{array}\right),
	\label{eq:singular-line-tension}
\end{equation}
respectively.
The solution curves of the parallel vectors operator for this simple model can then be derived in closed-form.
They are given by
\begin{equation}
	\mathbf{B}\|(\mathbf{B} \cdot \nabla) \mathbf{B}=\left\{\mathbf{x}_\pm\mid z\in\mathbb{R},\ z^2<\mu \right\},
\end{equation}
where
\begin{equation}
	\mathbf{x}_\pm=\left(\frac{B_0 z}{\sqrt{-\mu-z^2}},
		2\pm\sqrt{-\mu-z^2},z\right)^\top.
\end{equation}
The positive solution belongs to the vortex core line in the upper half and the negative solution to the bifurcation line in the bottom half of the domain $\Omega$.
An illustration of these solution curves with $\mu=1$ and $B_0=0.1$ is shown in \autoref{fig:singular-line-bifurcation} (left).

Due to their hyperbolic nature, bifurcation lines are associated with local surfaces of converging and diverging field lines in their vicinity.
Analogous to separatrices, these bounded two-dimensional manifolds locally separate the magnetic field, which identifies them with quasi-separatrix layers.
Conventional approaches to detecting QSLs rely on global field-line tracing.
In contrast, bifurcation lines provide a means to extract QSLs locally.
Each point \(\mathbf{x} \in S\) along the bifurcation line is displaced in the positive and negative directions of the two non-parallel eigenvectors \(\mathbf{v}_1\) and \(\mathbf{v}_3\) by a small offset \(\epsilon \ll 1\), i.e., $\mathbf{x}_{i,\pm} = \mathbf{x} \pm \epsilon \mathbf{v}_i$, followed by magnetic field-line integration.
The integration is performed in forward and backward directions, yielding the four local QSL surfaces of the X-line.
The bifurcation manifolds for the flux rope model \autoref{eq:solar-flux-rope} are shown in \autoref{fig:singular-line-bifurcation} (right). See Section~\ref{sec:harris-current-sheet} for application to a numerical plasma simulation.

\subsection{Filtering}
\label{sec:extraction-of-bifurcation-lines}
The Sujudi--Haimes criterion (i) is only a necessary condition for bifurcation lines or vortex core lines, so not all lines that satisfy this condition are such lines, or even valid magnetic field lines.
Hence, further filtering of the output of the parallel vectors operator is required.

Since we are only interested in field lines of the magnetic vector field, the angle between the vector field $\mathbf{B}$ and the tangent $\bm{\gamma}$ of the raw PV line
\begin{equation}
	\alpha(\bm{\gamma}, \mathbf{B})=\cos^{-1}\left(\frac{|\bm{\gamma}\cdot \mathbf{B|}}{\|\bm{\gamma}\|\|\mathbf{B}\|}\right)\le\tau_\alpha,
	\label{eq:feature-angle}
\end{equation}
is used filter for field line quality, where we set a threshold $\tau_\alpha$.
Only line segments with a low $\alpha$ are sufficiently tangent to a nearby field line.
Any line segment exceeding the user-defined threshold $\tau_\alpha$ is removed.
The value of the minimum possible angle $\alpha$ is constrained by the resolution of the simulation domain.
The higher the resolution of the simulation, the lower the angle threshold can be chosen (due to the finite-difference approximation of the tangent).
In a typical plasma simulation, a reasonable choice is $\tau_\alpha=15^\circ$.

To decide whether the line is a bifurcation line or a vortex core line, we can use the eigenvalues of the Jacobian matrix at each point on the line.
If all eigenvalues are real, i.e., condition (ii) from Section~\ref{sec:bifurcation-lines} is satisfied, the resulting manifold is a bifurcation line.
All other line segments can be discarded.

To ensure that criterion (iii) is satisfied, we filter the PV lines by the strength of its hyperbolic behavior, in the following referred to as \textit{feature strength} or \textit{hyperbolicity} of the bifurcation line.
We quantify the hyperbolicity by the product of the two non-parallel eigenvalues
\begin{equation}
	\phi(\mathbf{x}):=-\left(\lambda_1(\mathbf{x})\lambda_3(\mathbf{x})\right)\ge\tau_\phi,
	\label{eq:feature-strength}
\end{equation}
with a threshold $\tau_\phi$.
$\phi$ returns positive values only if there is a sign flip between $\lambda_1$ and $\lambda_3$.
Any point on the line with $\phi\ge\tau_\phi$ is filtered out.

The threshold $\tau_\phi$ of the hyperbolic strength has to be chosen according to the physical units of the system.
For example, a saddle point of the in-plane 2D vector field can be linearized and thus takes the shape $B(x)=(\lambda x, -\lambda y)^\top$, so we find $\phi=-(\lambda\cdot-\lambda)=\lambda^2$, which carries units of $T^2/m^2$.
We can get a non-dimensional $\phi$ by normalizing the units of the simulation to a natural field strength $B_0$ and length scale $L_0$ of the system.
Assume now that $\lambda=1$ is the strongest shear we expect in our normalized system, then $\phi=1^2=1$ will be the maximum possible feature strength.
If we want to find all lines up to half of the strongest shear $\lambda=0.5$, we would need to set our threshold to $\tau_\phi=(0.5)^2=0.25$.

For vortex core lines, we omit the regions where the imaginary part of the smallest eigenvalue of $\nabla\mathbf{B}$ is below a certain strength of rotation
\begin{equation}
	\psi=\min_{i=1,2,3}|\text{Im}(\lambda_i)|\le\tau_\psi,
	\label{eq:feature-strength-vrotex}
\end{equation}
where $\tau_\psi$ is the threshold.
Any points on the line with $\psi>\tau_\psi$ are filtered out.

\subsection{Quasi X-lines}
\label{sec:qxls}
The identification of X-lines as bifurcation lines with the parallel vectors operator works very well in smooth MHD simulations and with low magnetic guide fields.
However, the extraction is very sensitive to noise (such as the noise present in PIC simulations due to discrete number of particles per cell), curvature and high longitudinal components of the vector field.
This is because condition (iii) from Section~\ref{sec:bifurcation-lines} assumes that the two hyperbolic in-plane directions of the field along the line are the dominant directions (largest eigenvalues of the Jacobian $\mathbf{B}$).
However, this condition breaks down in the case of a high magnetic field strength longitudinal to the line, such as in guide field reconnection.
Actually, most of astrophysical magnetic reconnection is driven by strong guide fields.
For example, in typical 3D PIC simulations of turbulence, the guide field is usually four times larger than the magnetic field associated with the field gradient, such as found in current sheets~\cite{munozElectronInertiaEffects2023}.
Since such scenarios often lead to complex and turbulent dynamics, bifurcation lines can therefore be too restrictive and prone to errors, leading to insufficient solutions in difficult situations.

In the following, we will introduce \textit{quasi X-lines} or \textit{QXLs} as a relaxed version of bifurcation lines.
In the following we assume that the guide field dominates the flow, i.e., the angle between line-tangent $\gamma$ of the solution and the magnetic vector field exceeds the threshold $\tau_\alpha$ (\autoref{eq:feature-angle}), leading to very small lines or points.
However, we can still use these points to find the X-lines, hidden underneath the guide field.

Instead of using the filtered solutions of the parallel vectors operator, we search for potential points along the raw, unfiltered PV lines that satisfy the bifurcation line criterion (\autoref{eq:feature-strength}), i.e., we filter only by hyperbolicity and \textit{not} by angle (\autoref{eq:feature-angle}).
For each of these pre-filtered parallel vector lines, we then search for the point with the largest feature strength.
In the next step, we will use these points of largest hyperbolicity as seeds for field line integration of the magnetic vector field.
Since these points are also regions where the magnetic field aligns with its own tension, they can be identified as part of a bifurcation line.
This ensures that the quasi X-lines correspond to the X-lines of the magnetic vector field we search.

Let $S\subset\Omega$ be the set of all points of a single parallel vector line.
We then define the seed points for the QXL integration as
\begin{equation}
	\mathbf{\hat{x}}:=\argmax_{\mathbf{x}\in S}\phi(\mathbf{x}).
\end{equation}
The existence of these points can be seen in the following way: since a pre-filtered PV solution line $S$ can be parameterized by a closed interval $S(t):[0,1]\rightarrow \mathbb{R}^3$ with $t\in[0,1]$, and $\phi(S(t)):[0,1]\rightarrow \mathbb{R}$ is a scalar function along this interval, $\phi$ must have a global maximum.
On the other hand, this means that every initial PV line contains at least one seed point for the integration of the QXL\@.
If we find more than one point per line we choose one randomly.
We can choose any of them because, especially in degenerate cases, multiple points on the line can have similar hyperbolicity and each of them serves equally well as a seed for the QXL\@.
However, if already pre-filtered by $\phi$, most of the PV lines already have a single absolute maximum of hyperbolicity along the line.

Note, that we could also use all the points up to a certain hyperbolicity $\phi(\mathbf{x})$, which would give us multiple (in theory infinitely many) QXLs next to each other, and all of them show the desired in-plane hyperbolic behavior.
This can increase the statistics of consecutive calculations, e.g., for the reconnection rate estimation (See Section~\ref{sec:relative-reconnection-rate}), but to locate regions of high magnetic reconnection activity, using the maximum hyperbolicity point to seed single QXLs is mostly sufficient.

As a last step, the field lines need to be filtered with the same $\phi(\mathbf{x})$ as the initial PV lines.
This ensures strong hyperbolic behavior along all the field line segment.
The result is what we call the quasi X-line. The pseudocode for the quasi X-line extraction is shown in Algorithm~\ref{alg:qxl}. See Appendix~\ref{sec:validation} for further validation of QXLs as reconnection X-lines in guide field scenarios

Up to this point, the extraction of X-lines of the magnetic field as quasi X-lines required only the magnetic field itself, i.e., no other quantities, such as electric field or charge/current densities, were required.
However, we can use these quantities to further refine the solution lines by filtering for different derived signs of magnetic reconnection, e.g., non-ideal electric fields or reconnection rates.

\begin{figure}
	\begin{algorithm}[H]
		\caption{Extraction of Quasi X-Lines}
		\label{alg:qxl}
		\setstretch{1.5}
		\begin{algorithmic}[1]
			\State Compute $\nabla\mathbf{B}$ and $(\mathbf{B} \cdot \nabla) \mathbf{B}$ for all points $\mathbf{x}\in\Omega$
			\State Extract all lines $\mathbf{B}\|(\mathbf{B} \cdot \nabla) \mathbf{B} =\{S_1\cup S_2\cup \ldots\cup S_m\}\subset\Omega$
			\For{all parallel vector lines $S_i$ with $i\in\{1,\ldots,m\}$}
			\State Filter lines $S_i$ by $\phi$
			\State Find the point $\mathbf{\hat{x}}\in S_i$ with largest feature strength $\phi(\mathbf{x})$
			\State Add $\mathbf{\hat{x}}$ to the set of seed points $\hat{S}$
			\EndFor
			\State Start field lines from seed points $\mathbf{x}\in\hat{S}$
			\State Filter field lines by feature strength $\phi$
		\end{algorithmic}
	\end{algorithm}
\end{figure}

\subsection{Estimation of the reconnection rate}
\label{sec:relative-reconnection-rate}

A challenging part of determining the reconnection rate \autoref{eq:relative-reconnection-rate} of reconnection events in turbulence is the computation of the inflow magnetic field $B_{\text{in}}$ and charge density $\rho_{\text{in}}$, since we need to measure them just a few electron inertial lengths away from the X-line or the current sheet.
This can be done by shifting each point $\mathbf{x}$ of the X-line $S$ by a small $\delta$ in the inflow direction $\mathbf{n}$ normal to the current sheets.
Fortunately, the eigenvectors $\mathbf{v}$ of the Jacobian of the magnetic vector field $\nabla\mathbf{B}$ already provide a position- and field-dependent basis of the vector space relative to the current sheet.
By shifting in the direction of the eigenvector corresponding to the largest eigenvalue $\mathbf{v}_1$, and measuring $\mathbf{B}$ and $\rho$ at the shifted line points, we can finally estimate the inflow magnetic field strength $B_{\text{in}}$ and $\rho_{\text{in}}$ as the average over our samples
\begin{equation}
	B_{\text{in}}\approx \frac{1}{2n_S}\left(\sum_{\mathbf{x}\in S}\mathbf{B}(\mathbf{x}+\delta\mathbf{n}) + \mathbf{B}(\mathbf{x}-\delta\mathbf{n})\right)
\end{equation}
and
\begin{equation}
	\rho_{\text{in}}\approx \frac{1}{2n_S}\left(\sum_{\mathbf{x}\in S}\rho(\mathbf{x}+\delta\mathbf{n}) + \rho(\mathbf{x}-\delta\mathbf{n})\right),
\end{equation}
where $n_S$ is the number of points on the X-line $S$ and $\mathbf{n}=\mathbf{v}_1/\|\mathbf{v}_1\|$.
The shift $\delta$ is typically chosen to be a few times the electron inertial length.
Additionally, the current density $\mathbf{J}$ can be evaluated at $\mathbf{x}\pm\delta\mathbf{n}$ to verify that $B_{\text{in}}$ and $\rho_{\text{in}}$ are actually measured outside the current sheet.
In case of two neighboring current sheets, $\mathbf{J}$ can also be used to avoid incorrect normalization in \autoref{eq:relative-reconnection-rate}.

The line integral in \autoref{eq:reconnection-rate} can be estimated straight forward, e.g., by a discrete sum using the trapezoidal rule
\begin{equation}
	\int_{\text{X-line}}E_\|\mathrm{d}s\approx\frac{1}{2}\sum_{i=0}^n (E_\|^{i+1}+E_\|^i)\Delta s_i
\end{equation}
with
\begin{equation}
	E_\|^i=\frac{\mathbf{E}(\mathbf{x}_i)\cdot\mathbf{B}(\mathbf{x}_i)}{\|\mathbf{B}(\mathbf{x}_i)\|},
\end{equation}
where we sum over line segments $\Delta s_i=\|\mathbf{x}_{i+1}-\mathbf{x}_i\|$.

The benefit of calculating the reconnection rate this way is, again, full locality.
No need to assume a global inflow direction.
By using the eigenvectors of the Jacobian along the X-line, the computation handles curvature naturally and even in presence of strong guide fields or asymmetric reconnection, the algorithm delivers reliable estimates for the reconnection rate.
But note that if the input data is noisy, like in some of the data to be shown, this calculation should be handled carefully.
Numerical estimation of the Jacobian matrix by finite-differences and its eigen-decomposition is highly sensitive to noise and can lead to eigenvectors with varying signs (since $-\mathbf{v}$ solves the eigenvalue equation equally well as $\mathbf{v}$) which should be aligned before shifting the X-lines.

\subsection{Shear layers}
\label{sec:shear-layers}
In the context of fluid dynamics, a shear layer is a thin region of high shear stress of the flow vector field.
\citet{schafhitzelVisualizingEvolutionInteraction2011} introduced the extraction of such shear layers as a local feature of vector fields.
Similar to the $\lambda_2$ criterion for extraction of vortex regions~\cite{schafhitzelTopologyPreservingBasedVortex2008}, the shear layer is defined for a vector field $\mathbf{B}$ using the decomposition of the Jacobian $\nabla\mathbf{B}$ into symmetric and antisymmetric parts, i.e.,
\begin{equation}
	\nabla \mathbf{B}= \mathbf{S} + \boldsymbol\Omega
\end{equation}
with
\begin{equation}
	\mathbf{S}=\frac{(\nabla \mathbf{B})+(\nabla \mathbf{B})^\top}{2}\quad\text{and}\quad\boldsymbol\Omega=\frac{(\nabla \mathbf{B})-(\nabla \mathbf{B})^\top}{2}.
\end{equation}
The antisymmetric part $\boldsymbol\Omega$ is called the rotation tensor and is used to define the $\lambda_2$ value for vortex extraction as the second-largest eigenvalue of $\boldsymbol\Omega$.
The symmetric part $\mathbf{S}$ is called the shear strain tensor and encodes the linearized spatial shear of the local vector field.
This shear tensor can be used to define a single scalar quantity measuring shear at any given point of the vector field.
The so-called $I_2$ value is defined as the (negative) second invariant of the shear strain tensor, i.e.,
\begin{equation}
	I_2:=-\frac{1}{2}\left(\text{tr}(\mathbf{S}^2)+\text{tr}(\mathbf{S})^2\right)=-\left(\lambda_1\lambda_2+\lambda_1\lambda_3+\lambda_2\lambda_3\right),
	\label{eq:i2-value}
\end{equation}
where we used the eigenvalues $\lambda_i$ for $i\in\{1,2,3\}$ of $\mathbf{S}$.
In principle, the $I_2$ value can be visualized by direct scalar field visualization techniques like volume rendering or isocontouring, where positive and large values of $I_2$ represent higher shear stress.
\textit{Shear layers} are defined as the area enclosed by $I_2=0$ isosurfaces.

As a derived quantity of the symmetric part of the Jacobian $\nabla\mathbf{B}$, the $I_2$ value quantifies the shear level within the vector field, deliberately omitting any rotational influences.
On the other hand, diagnostics for current sheets using the current density $\|\mathbf{J}\|\sim\|\nabla\times\mathbf{B}\|$ also measure the rotational aspects of the magnetic field.
This reveals the advantage of employing the $I_2$ value over the current density for locating current sheets: it specifically targets shear without conflicting it with currents arising from magnetic field rotation, such as is present, e.g., in plasmoids.
By design, the $I_2$ value focuses on the symmetric part of the field, reflecting purely sheared fields.

\section{Implementation and visualization}
\label{sec:implementation}
Most of the algorithms presented in this work are implemented as ParaView\cite{squillacoteParaViewGuideParallel2007} filter plugins and are built around the Visualization Toolkit\cite{schroederVisualizationToolkitObjectoriented2006} (VTK).
Developed by Kitware, VTK is one of the most widely used open-source libraries for efficient algorithms that operate on arbitrary grid and geometric data, both in scientific research and industrial applications.
It provides highly optimized data structures for processing 3D vector fields, making it well-suited for the extraction and subsequent visualization of magnetic reconnection phenomena in three-dimensional plasma simulations.

VTK already provides a reference implementation of the parallel vectors operator, called \textit{vtkParallelVectors}.
The algorithm works as follows:
\begin{enumerate}[(i)]
	\item Each face of each cell of the simulation domain gets triangulated.
	\item On each face triangle \autoref{eq:pv-criterion} is analytically solved for the point where the two linearly interpolated vector fields are parallel (using barycentric interpolation).
	\item At most two such points per cell are kept and used to create a line segment crossing the cell.
	\item Adjacent links between cells are stitched into a \textit{vtkPolyLine}.
	\item Polylines are finally written into a \textit{vtkPolyData} object, whose points come from a merge-locator (so duplicates are collapsed) and whose lines are produced by a small in-memory graph builder \textit{vtkPolyLineBuilder}.
\end{enumerate}
This produces the \textit{raw} parallel vector lines.
The main part of the algorithm is parallelized using multi-threading, since the calculations for each cell are independent of each other.

The filter class further exposes virtual hooks, so subclasses can (a) skip certain faces, (b) add extra ``criterion'' scalars to each output point or (c) post-process the final polydata.
We use (b) to implement the angle criterion (\autoref{eq:feature-angle}), as well as the bifurcation lines feature strength (\autoref{eq:feature-strength}) and (c) to filter out weak solutions.

\section{Results}
\label{chap:results}
In this section, we demonstrate the application of our method to three distinct plasma simulation setups, chosen to span a range of physical regimes and complexities.
First, we consider a fully kinetic simulation of a Harris current sheet, a classical benchmark for magnetic reconnection studies.
Second, we analyze a magnetohydrodynamic simulation of a coronal flux rope eruption, representing a realistic reconnection scenario in the solar corona.
Finally, we apply our method to a hybrid-kinetic fluid simulation of a turbulent plasma, capturing time-dependent and inherently three-dimensional reconnection characteristics of the solar wind.
\subsection{Fully-kinetic model of Harris current sheet}
\label{sec:harris-current-sheet}
The Harris current sheet is a standard test case for magnetic reconnection.
It was found by Harris~\cite{harrisPlasmaSheathSeparating1962} as an exact equilibrium solution of the Vlasov--Maxwell equations.
The plasma pressure is constant along the $x$ and $z$-direction and balanced by the magnetic pressure, as the plasma is in equilibrium.
The guide field is set to $B_z=0.1B_x$, i.e., small compared to the transversal component.
To start the simulation, a small initial perturbation is added to the magnetic field in the middle of the $x$-$y$-plane, which triggers the magnetic reconnection process.
The simulation is then run for a certain time until the energy stored in the magnetic field gradient has dissipated.

The Harris sheet simulation was performed with the
fully kinetic Particle-in-Cell (PIC) code ACRONYM \citep{kilianInfluenceMassRatio2012}.
The simulation domain is sampled on a regular grid with a size of $N_x\times N_y\times N_z=64\times 128\times 128$ grid cells and periodic boundary conditions.
The grid cell size is chosen to be one Debye-length due to the stability requirements of explicit fully kinetic PIC codes.
This implies that the domain is relatively small, mainly covering electron kinetic scales.
For typical solar corona densities ($\sim8\times10^{9}$ particles~${\rm /cm}^3$), this represents
a physical domain size of $(64\times128\times128)$~${\rm cm}^3$,
while for typical solar wind densities ($\sim 1$ particles~${\rm /cm}^3$),
the physical domain is instead $(60\times120\times120)$ ${\rm km}^3$.
The simulation was run for $t=100$ time steps,
which was sufficiently long to observe the onset of reconnection
due to the available magnetic flux in the simulation domain.
The time step was also chosen to satisfy the stability requirements of explicit fully kinetic PIC codes, i.e., the CFL condition for light waves.

\begin{figure}
	\centering
	\includegraphics[width=0.99\linewidth]{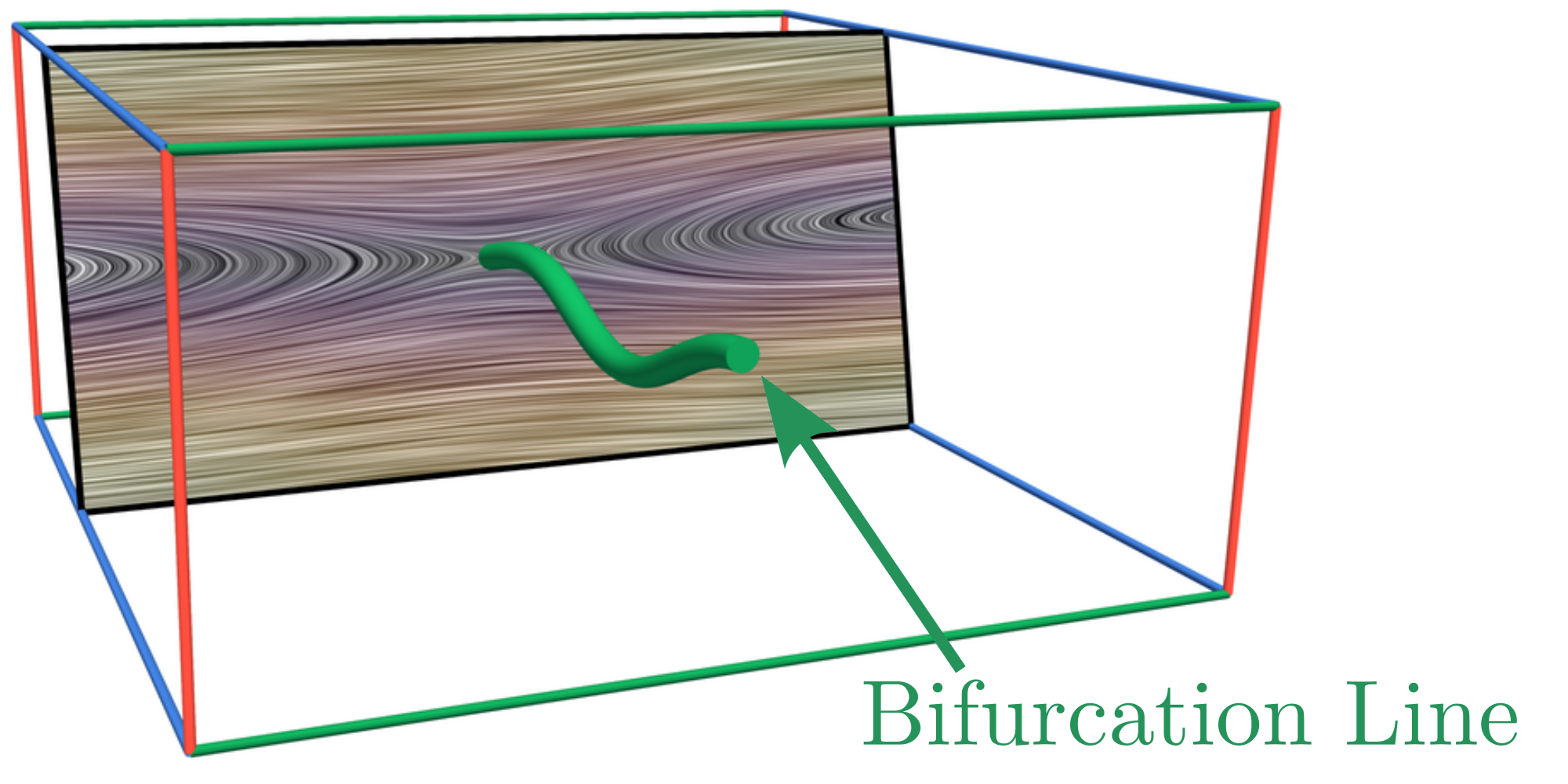}\\
	\includegraphics[width=0.99\linewidth]{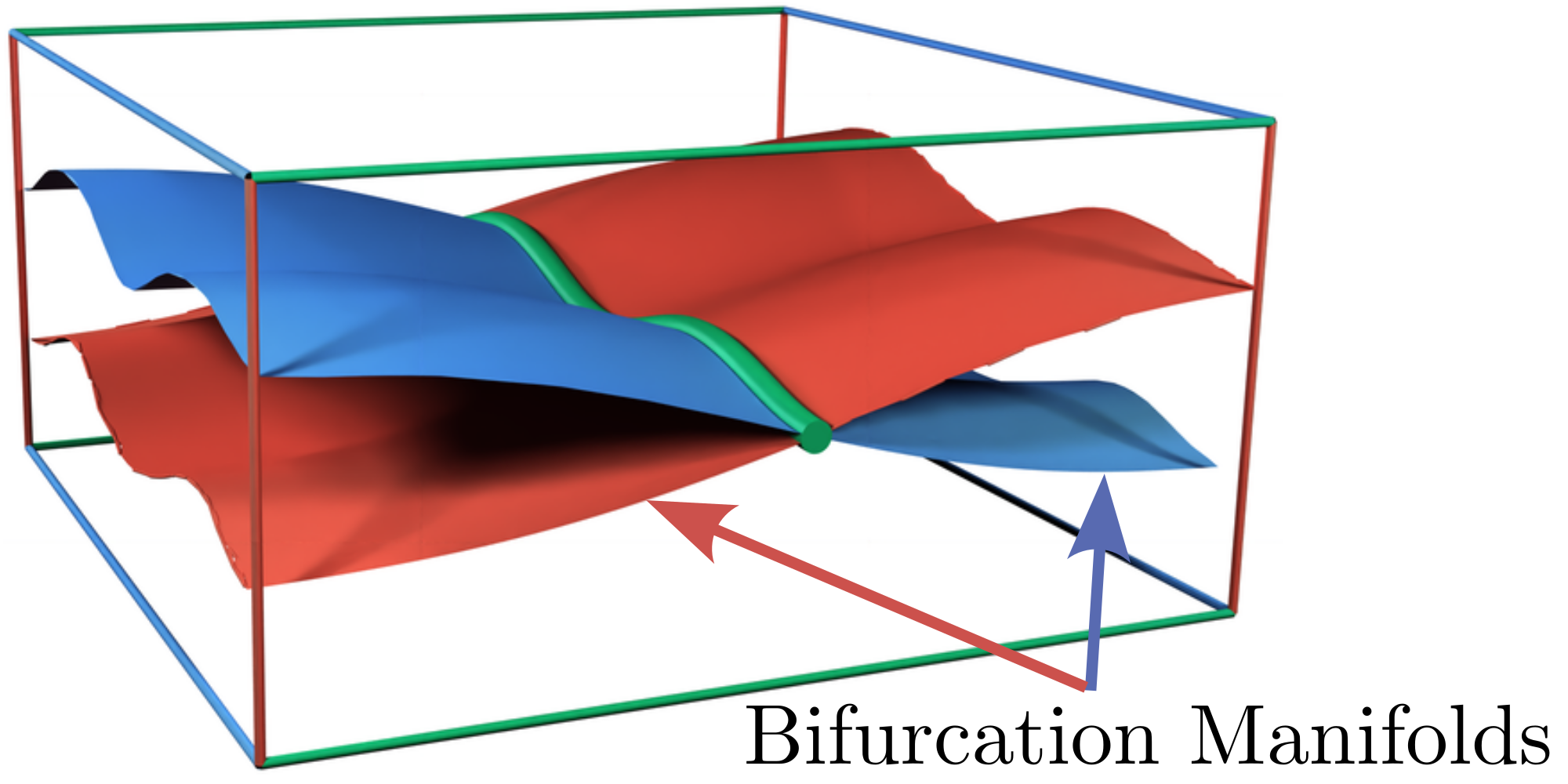}
	\caption{(Top) Harris sheet simulation with bifurcation line (green).
		The in-plane magnetic field is visualized using LIC.
		(Bottom) Bifurcation manifolds of converging (blue) and diverging (red) field lines.}
	\label{fig:harris-sheet}
\end{figure}

The main feature of this simulation is the formation of a strong current sheet in the $y$-direction by definition.
The current sheet is unstable, so after adding an initial perturbation, a very obvious magnetic reconnection topology occurs, driven by the tearing instability~\cite{zannaIdealTearingMode2016}.
In the middle of the $x$-$y$-plane, a clear X-line starts to emerge.
This X-line is accompanied by the formation of magnetic islands, which are regions of closed-loop magnetic fields.
The magnetic islands are separated by QSLs.
As the guide field of this simulation is small, secondary instabilities, such as the kink instability~\cite{zhuTearingInstabilityFlux1996}, occur.
These create large-scale waves in the plasma, which make the localization of X-lines especially challenging.

We can observe that the simple bifurcation line in this simulation is indeed an X-line.
This is because the raw parallel vectors line is parallel to the magnetic field, i.e., the feature angle $\alpha$ is small (See Section~\ref{sec:bifurcation-lines}).
In the plane perpendicular to the bifurcation line, the magnetic field shows hyperbolic behavior, i.e., we find an X-point of the magnetic field in the intersection of the plane with the bifurcation line.

As described in Section~\ref{sec:parallel-vectors-operator}, QSLs can now be extracted as bifurcation manifolds of converging and diverging field lines.
The resulting surfaces are shown on the right side of \autoref{fig:harris-sheet}.
We find that the bifurcation manifolds indeed separate the magnetic islands, which indicates that bifurcation manifolds, with their converging and diverging field lines, can be identified with QSLs.

However, this is a simplified scenario, where the magnetic field is very well-behaved, and the X-line is very obvious.
In more realistic or turbulent scenarios, the magnetic field is not as laminar and the X-lines are not as obvious.

\subsection{MHD model of coronal flux rope eruption}
\label{sec:coronal-flux-rope-eruption}
As a next example, we demonstrate identification of magnetic reconnection in a coronal flux rope simulation.
As mentioned in the Introduction~\ref{sec:introduction}, this is an explosive phenomenon, associated to solar flares and powered by magnetic reconnection.

The identification of reconnection in simulations of this process, also often linked to turbulence, is very challenging due to the complexity and fast time dependence
of the relevant magnetic field topology.
It also includes twisted magnetic field lines, but, different from our previous application (Section \ref{sec:harris-current-sheet}), reconnection here is mainly characterized by a small to negligible guide field.
This means that reconnection occurs here also in true null points, an important consideration for the choice of identification algorithm.

The magnetic field used for our analysis is from a data-driven simulation of a coronal flux rope eruption, which was recently analyzed and visualized with a local approach by \citet{wangMethodDeterminingLocations2023}.
The simulation was performed by \citet{guoThermodynamicMagneticTopology2023} using the full MHD equations to solve for a physically realistic scenario.
It is initialized with a magnetic flux rope similar to observations and further constrained by observed magnetic and velocity fields.
The authors showed that the simulation can reproduce the observed characteristics of the X1.0 flare on 2021 October 28.
To enable a qualitative comparison of our results with the extraction of magnetic reconnection structures using existing methods, we will use the same time step as Wang et al., which is at 15:32 UT.\@
The simulation domain is sampled on a regular grid with a size of $N_x\times N_y\times N_z=330\times 260\times 280$ grid cells.
This represents a physical domain of $(470\times372\times399)$~${\rm Mm}^3$.

\begin{figure}
	\centering
	\includegraphics[width=0.99\linewidth]{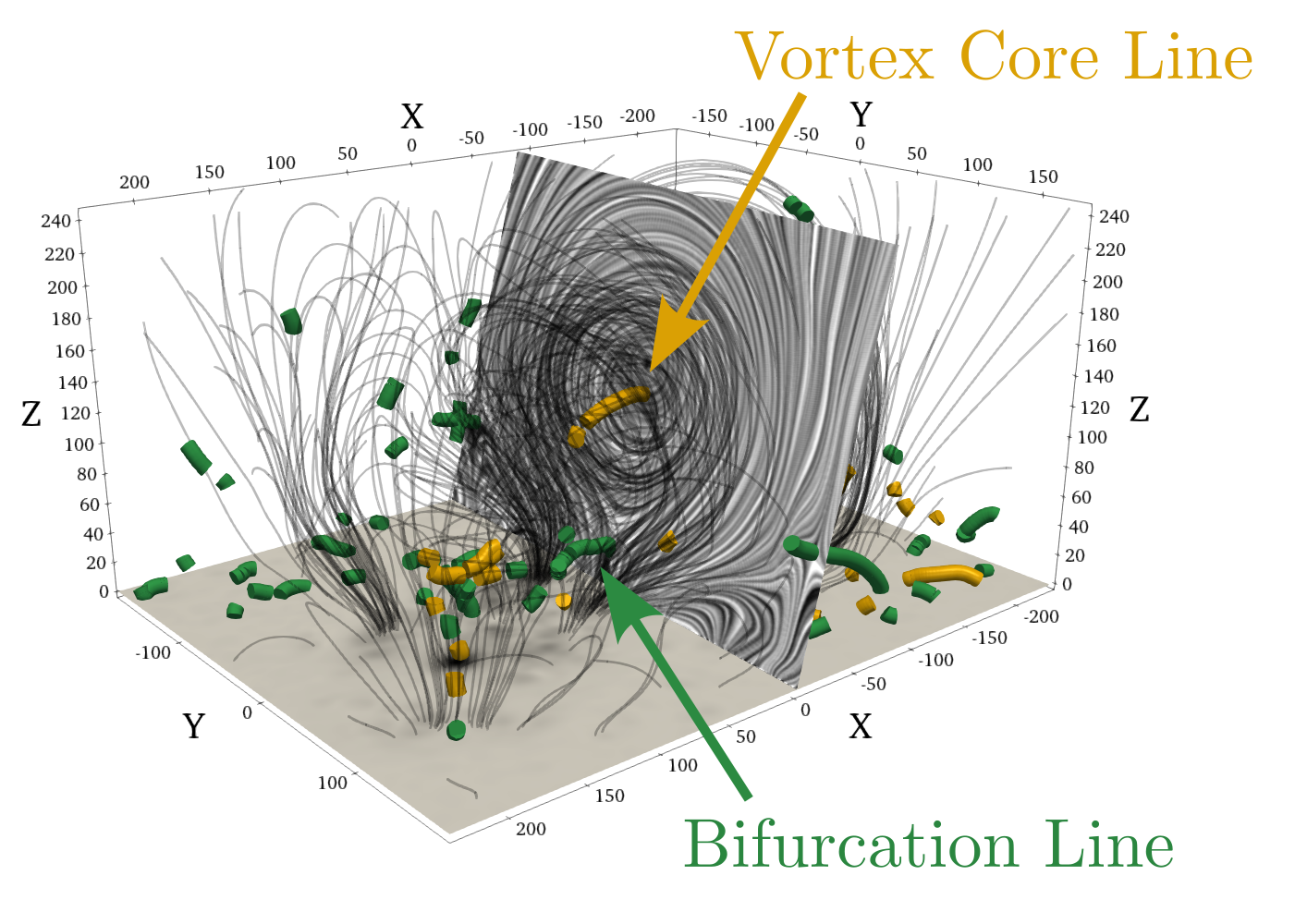}\\
	\includegraphics[width=0.7\linewidth]{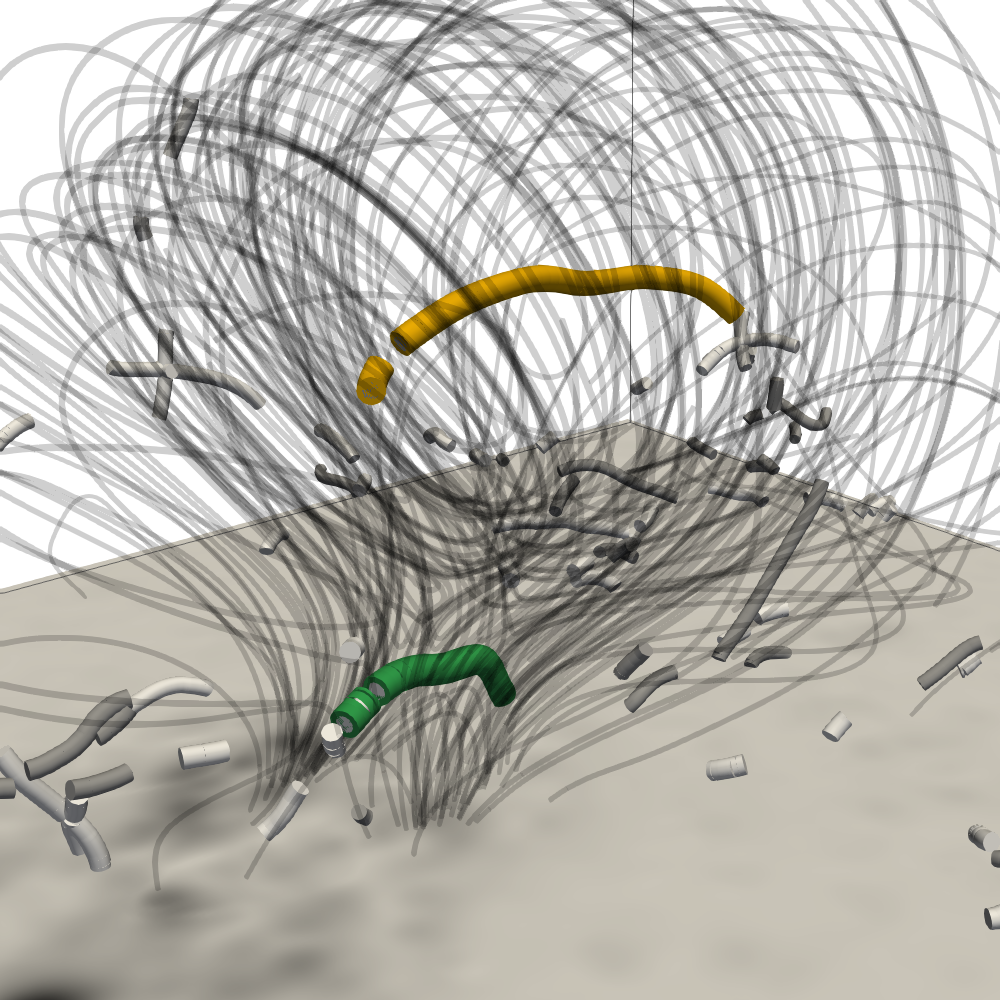}
	\caption{(Top) Visualization of the coronal flux rope simulation with bifurcation lines (green) and vortex core lines (yellow).
	The magnetic field strength of the solar corona (the $z=0$ plane) is shown as a heat map in grayscale. (Buttom) Zoom into the relevant structure of the prominence showing a typical X-line and O-line configuration.}
	\label{fig:flux-rope}
\end{figure}

The main feature of this simulation is the formation of a flux rope, a set of twisted magnetic field lines, emerging from the surface of the sun.
Similar to the simple flux rope model discussed in Section~\ref{sec:parallel-vectors-operator}, an X-line is expected in the shear zone between the flux rope and the surface of the sun (the $z=0$ plane), driving the magnetic reconnection of the eruption.
The extraction of X-lines as bifurcation lines (green) is shown in \autoref{fig:flux-rope}.
Vortex core lines (yellow) mark the center of the flux rope.

The field lines of the magnetic vector field of this simulation shows the typical structure of a coronal flux rope.
The points of high O-type strength found by Wang \textit{et al.} are well captured by the vortex core lines, which lay at the center of the flux rope.
However, the approach of Wang \textit{et al.} was not able to find the X-lines below the flux rope, which we have extracted as bifurcation lines with the parallel vectors operator.
At these X-lines, the highest rate of reconnection is expected.
Further, all features extracted with our approaches are solely based on the magnetic field, while for the approach of Wang \textit{et al.} the electric field is also necessary. However, the electric field is not always available in simulations and observations.
Note that in those MHD simulations, the electric field is directly proportional to the current density via the resistive MHD Ohm's law,
which in turn can be determined by the curl of the magnetic field.
So in principle only the magnetic field is necessary to determine the electric field.
But this proportionality only works in resistive MHD, whereas in other plasma models the relationship can be so complex that it cannot be directly calculated in a practical way.

\subsection{Hybrid-kinetic plasma turbulence in the solar wind}
\label{sec:solar-wind}
The third and last application of our method is to a simulation
of plasma turbulence for solar wind conditions.
The solar wind is a very collisionless and turbulent plasma
escaping from the solar corona in the interplanetary medium.
The turbulence in this plasma is believed to be heavily influenced by magnetic reconnection~\cite{munozElectronInertiaEffects2023, masaakiyamadaMagneticReconnection2022}.

There are not many direct measurements of  reconnection embedded in turbulence in the solar wind itself, since reconnection
is more often observed at very large scales \citep{Gosling2012, Wang2023a}, but such scenarios are backed by observational evidence in the turbulent Earth's magnetosheath by the MMS space mission~\cite{vorosMMSObservationMagnetic2017}, where magnetic reconnection has often been observed even at very small scales.
The type of reconnection that is observed in this turbulent environment differs to that of our first application in that it is mainly guide field reconnection, since the solar wind is permeated by the magnetic field originated from the Sun, the so-called Parker spiral~\cite{parkerDynamicsInterplanetaryGas1958}.

The solar wind simulation was performed with the CHIEF code~\cite{munozNewHybridCode2018}.
CHIEF is a hybrid kinetic PIC code, where the ions are treated as kinetic particles and the electrons as a fluid.
The simulation domain is sampled on a regular grid with a size of $N_x\times N_y\times N_z=256\times 256\times 256$ grid cells and periodic boundary conditions.
For typical solar wind densities ($\sim 1$ particles~${\rm cm}^3$),
this represents a physical domain of $(1280\times1280\times1280)$~${\rm km}^3$.
The physical setup and further discussions of this simulation are given by~\citet{munozElectronInertiaEffects2023}.

The field is chosen such that the initial turbulence level of the magnetic field fluctuations is similar to observations at the Earth's orbit or one astronomical unit from the Sun, i.e, $|\delta B/B|\sim0.24$ \citep{Bale2009}.
This leads later to a typical guide field of $\approx 1/4$ of the reconnecting magnetic field component in the current sheets formed out of turbulence, i.e., a very high longitudinal component.
The simulation was run for $t=114$ time outputs, where one time output occurs every 500 time steps of the simulation, and it is equivalent to $2 \Omega_{ci}^{-1}$, where $\Omega_{ci}$ is the ion cyclotron frequency.

\begin{figure}
	\centering
	\includegraphics[width=0.99\linewidth]{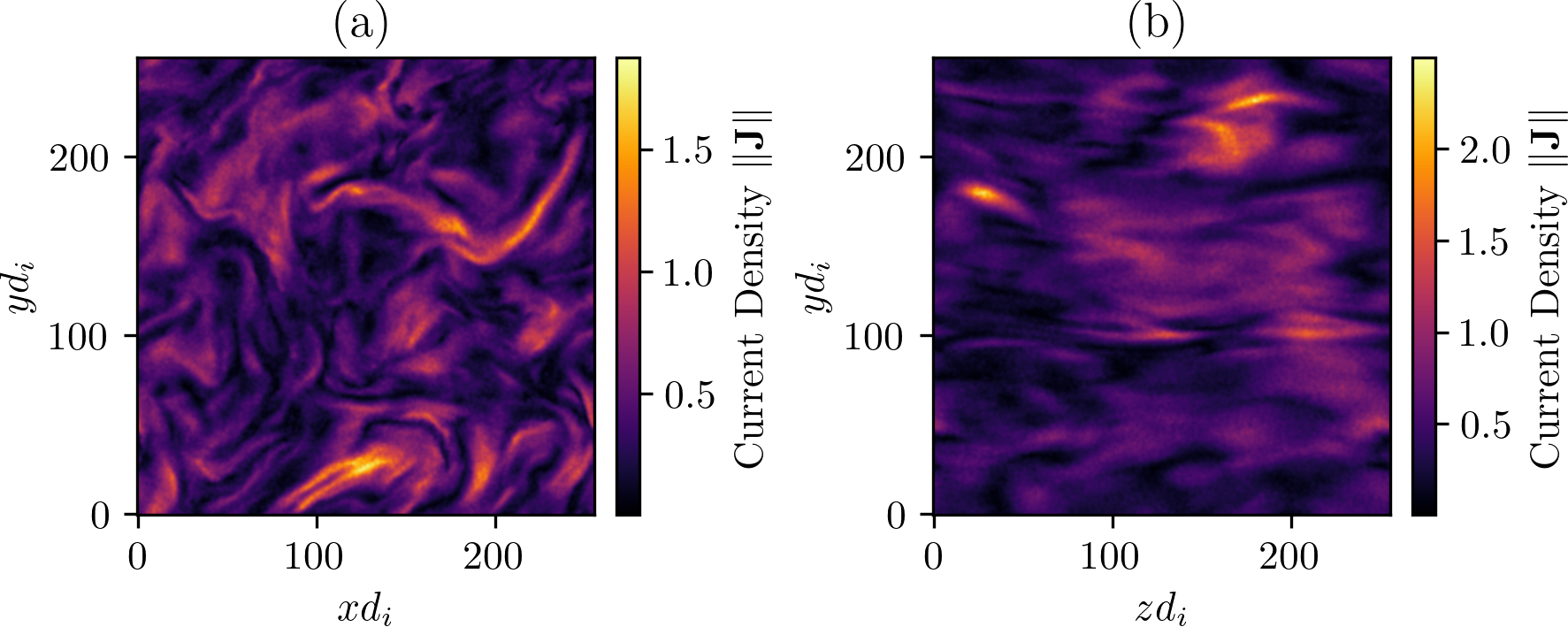}
	\caption{Hybrid-kinetic turbulence simulation at time $t=140 \Omega_{ci}^{-1}$ showing the current density $\|\mathbf{J}\|$ in (a) the $x$-$y$-plane at $z=0$ and (b) the $y$-$z$-plane at $x=0$. The magnetic guide field is directed along the $z$-axis.}
	\label{fig:solar-wind-old}
\end{figure}

The main feature of this simulation is the formation of multiple strong current sheets, with a corresponding high rate of magnetic reconnection events.
These occur due to the energy cascade from large scale (the initial perturbation) to small scales.
The occurrence of such current sheets is strongly coupled to magnetic reconnection.
The traditional approach to the localization of current sheets is the current density (dominated by the electron contribution), which is illustrated in \autoref{fig:solar-wind-old} by volume rendering.

The simulation is a challenging benchmark for the extraction of magnetic reconnection, because the turbulent nature of the turbulent plasma creates a very complex and highly fluctuating magnetic field.
The magnetic reconnection structures are not as well-behaved as in the simple Harris sheet simulation
that we used to validate our approach in Section~\ref{sec:harris-current-sheet}
or in the coronal flux rope simulation from Section~\ref{sec:coronal-flux-rope-eruption}, so the X-lines can potentially be very short and curved.
In addition, there is numerical noise because of the Lagrangian nature of the hybrid PIC simulation, leading to a noisy magnetic field and possibly spurious reconnection events.
\begin{figure*}
	\centering
	\begin{tabular}{ccc}
		\includegraphics[width=0.99\linewidth]{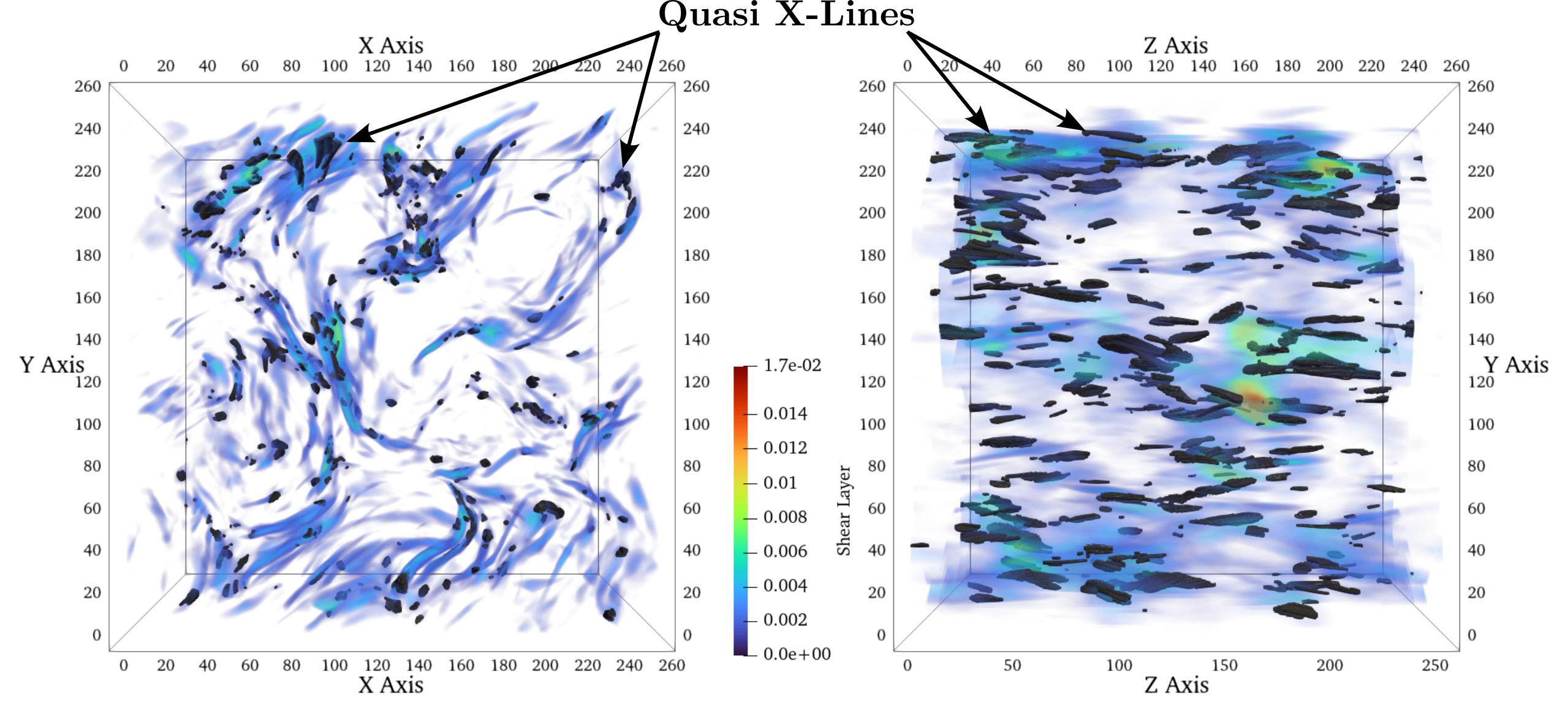}
	\end{tabular}
	\caption{Hybrid-kinetic turbulence simulation at time $t=140\Omega^{-1}_{ci}$ with quasi X-lines (black) and shear layers as volume rendered $I_2$ value.
		The magnetic guide field is directed along the $z$-axis.
		(Left) Front view, (Right) side view}
	\label{fig:solar-wind}
\end{figure*}

We introduced quasi X-lines as a relaxed version of bifurcation lines in Section~\ref{sec:qxls}, and developed an algorithm based on the parallel vectors operator that yields a more useful solution to the search for X-lines in turbulent guide field scenarios.
The result of the algorithm for magnetic reconnection in the solar wind simulation is illustrated in \autoref{fig:solar-wind}.

In addition to the localization of X-lines, we identified shear layers within the magnetic field using the $I_2$ value, a method effective for the extraction and visualization of current sheets in turbulent plasma.
To qualitatively analyze the spatial correlation between QXLs and these shear layers, in \autoref{fig:solar-wind} we overlay the QXLs with volume rendered shear layers using a linearly increasing transfer function.

From these visualizations, we can observe that the QXLs and the shear layer of the magnetic field correlate in space, i.e., the QXLs are located predominantly in the shear layers of the magnetic field.
This supports the conclusion that QXLs are indeed magnetic X-lines and that strong magnetic shear is a good indication for the occurrence of unstable current sheets, breaking into X-lines in the process of magnetic reconnection.

\begin{figure*}
	\centering
	\begin{tabular}{ccc}
		(a) $\mathbf{B}$  & (b) $\|\mathbf{J}\|$  & (c) $I_2$\\
		\includegraphics[width=0.27\linewidth]{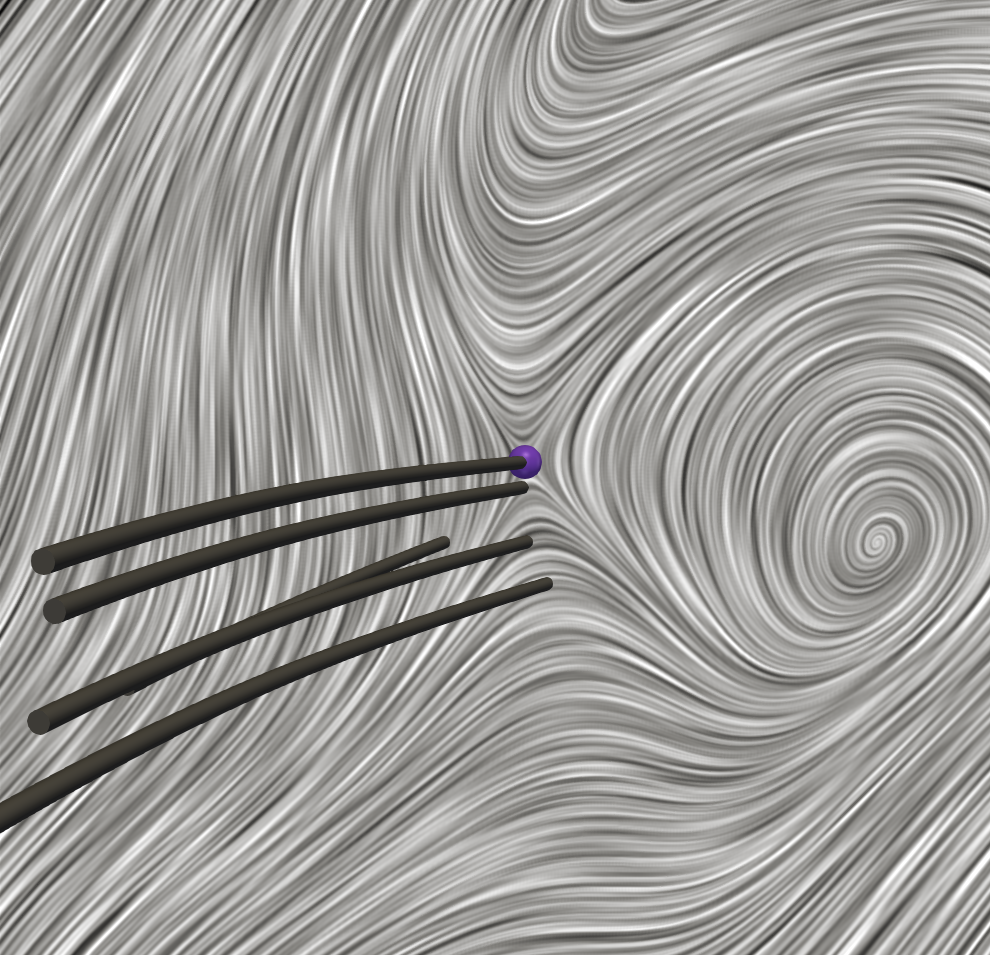}              &
		\includegraphics[width=0.3\linewidth]{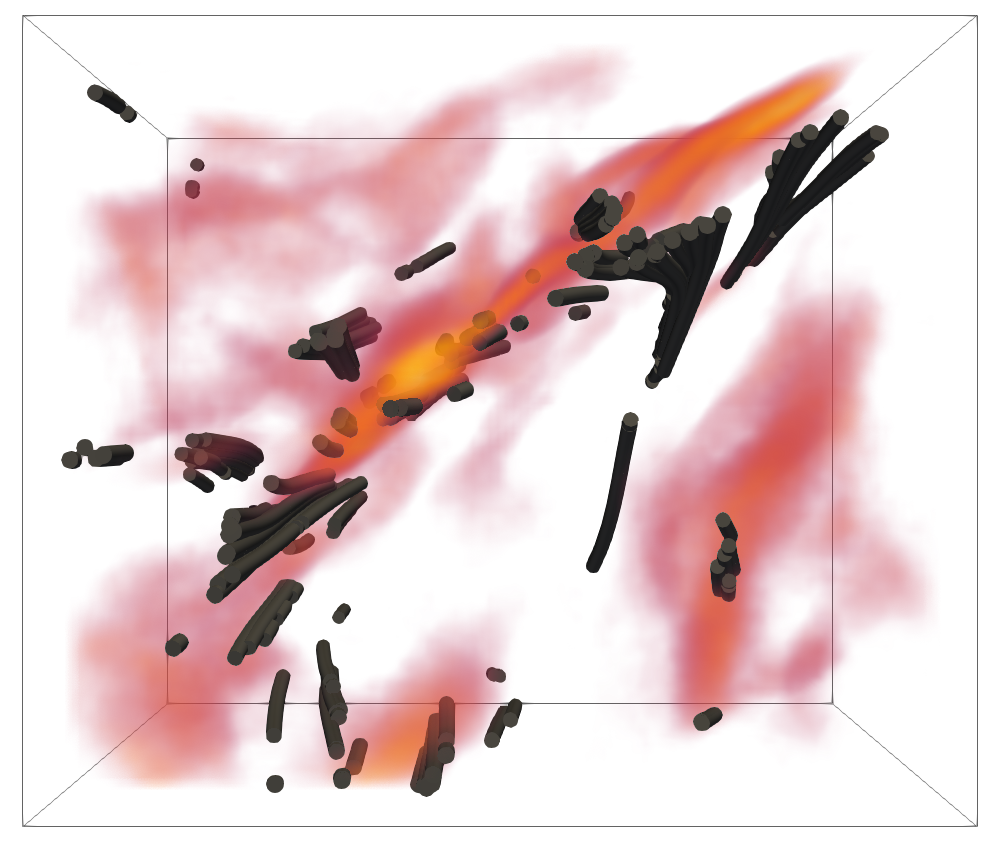} &
		\includegraphics[width=0.3\linewidth]{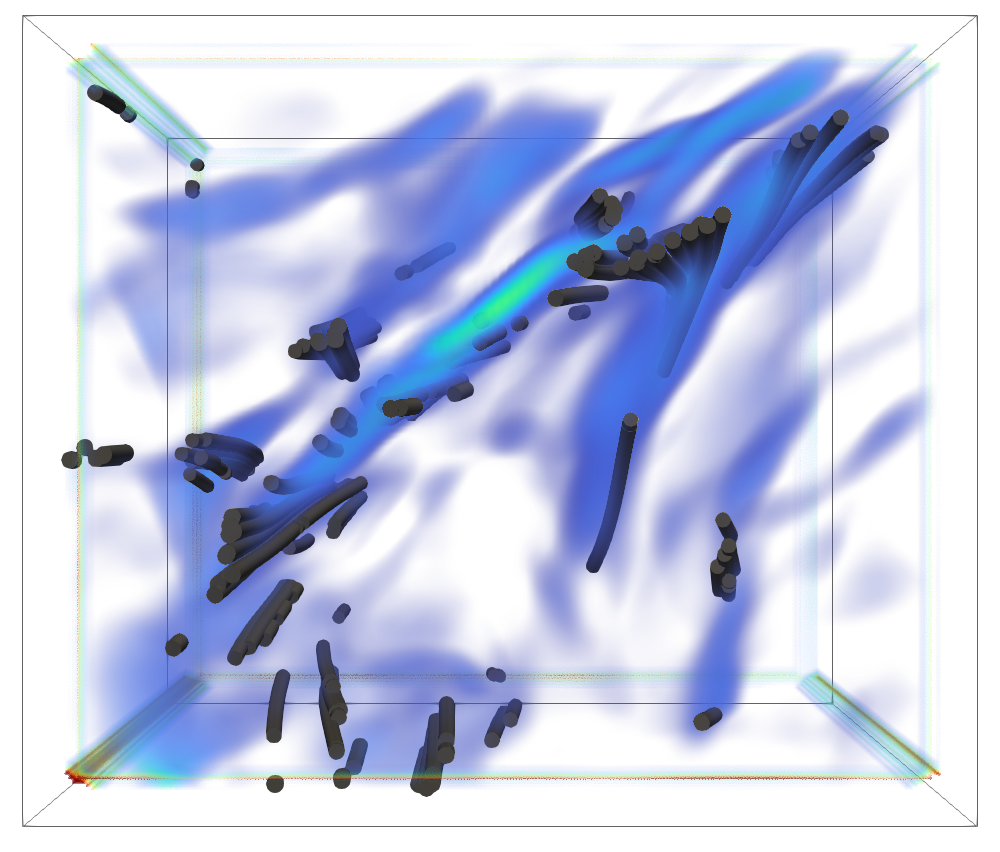}
	\end{tabular}
	\caption{Zoomed-in view of selected features in the hybrid-kinetic solar wind simulation.
		(a) Quasi X-line showing X-type null point (violet sphere) in the magnetic field in the plane perpendicular to the line tangent.
		(b) Quasi X-lines with volume rendered current density $\|\mathbf{J}\|$.
		(c) Quasi X-lines with volume rendered $I_2$ shear layers.}
	\label{fig:solar-wind-near}
\end{figure*}

In \autoref{fig:solar-wind-near}, we show zoomed-in views of the QXL extraction.
In the first plot (a) we show the in-plane magnetic vector field, perpendicular to the QXL tangent.
As can be seen, the line shows the expected X-type null point of $\mathbf{B}$ in this projection.
This is true for the whole length of the QXL.\@
Note that the neighboring lines do not show an X-point in this plane.
To see their X-points, the plane must be set perpendicular to their tangent.
In \autoref{fig:solar-wind-near} (b) and (c) we compare the location of the QXLs relative to the current density $\|\mathbf{J}\|$ and the $I_2$ shear layers, respectively.
We can see that in both cases the X-lines are located inside both, the current sheets and regions of high magnetic shear.
However, the $I_2$ value contains more X-lines than the $\|\mathbf{J}\|$.
This illustrates that both are similar, but not the same quantities.
If only $\mathbf{B}$ is available, the $I_2$ value can reliably locate current sheets as strong shear in the magnetic vector field, but they can also serve as a complementary tool for the analysis of magnetic reconnection.

\subsection{Temporal onset of magnetic reconnection}
\label{sec:temporal-onset-of-magnetic-reconnection}
\begin{figure}
	\centering
	\includegraphics[width=0.99\linewidth]{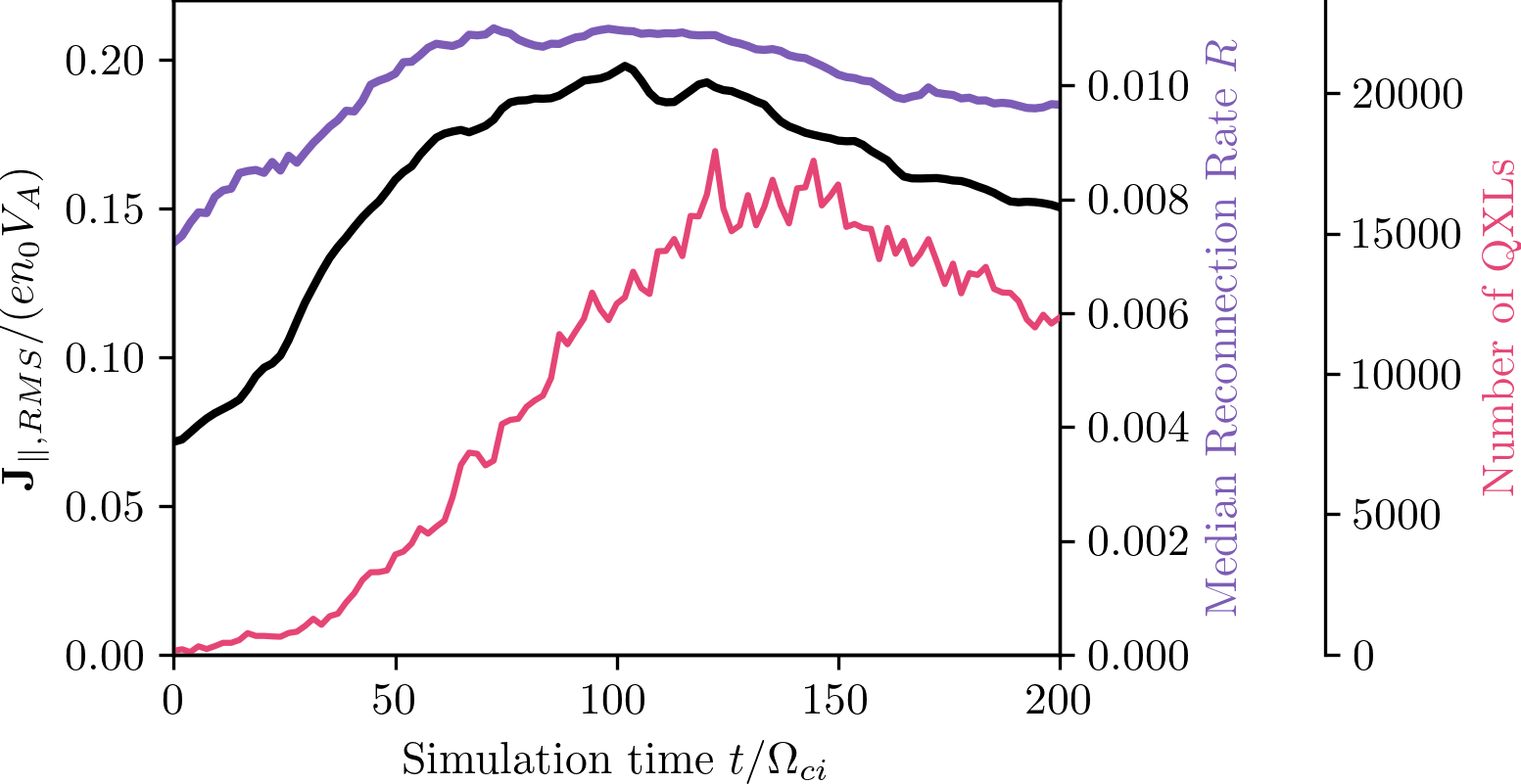}
	\caption{Comparison of the RMS value of the $J_{\|,RMS}\propto J_z$ component of the current density (black) with the median reconnection rate $R$ (violet) and the number of extracted quasi X-lines  (magenta).
		The reconnection rate has been smoothed using an exponential moving average with $\alpha=0.1$.
	}
	\label{fig:time-plot}
\end{figure}

The standard diagnostic to probe for the temporal onset of magnetic reconnection in plasma turbulence simulations is the strength of the current density $\mathbf{J}$ in the whole simulation domain, which is available from the simulation and shown in \autoref{fig:solar-wind}.
As discussed in the Introduction~\ref{sec:introduction}, the magnitude of the current density $\|\mathbf{J}\|$ is a good indicator for magnetic reconnection in turbulence simulations, as high values of current density imply thinner current sheets (steeper magnetic field gradient), leading to larger tearing mode growth rates and thus stronger magnetic reconnection.
\citet{munozElectronInertiaEffects2023} measured the root-mean-squared (RMS) current density $J_{\|,RMS}\propto J_z$ of the solar wind simulation in the plane perpendicular to the guide field, i.e., in $z$-direction, as a function of time. The evolution of the RMS current density is shown in \autoref{fig:time-plot} (black line).
The magnitude of $J_{\|,RMS}$ shows a peak near the time step $100\Omega_{ci}^{-1}$, while it was observed that magnetic reconnection starts to break the current sheets at time steps later than 40$\Omega_{ci}^{-1}$ to 60 $\Omega_{ci}^{-1}$.

Provided the location of QXLs, we can now also evaluate the reconnection rate as a function of time using our methods from Section~\ref{sec:relative-reconnection-rate} (violet curve in \autoref{fig:time-plot}). The calculation of the reconnection rate is described in Section~\ref{sec:relative-reconnection-rate} and the results for this simulation (distribution of reconnection rates) for selected times are summarized in the following Section~\ref{sec:estimation-reconnection-rate}.
We use the median over the set of all X-lines here because degenerate cases and numerical constraints can produce outliers with unphysically high reconnection rates.

By further counting the number of extracted QXLs, we can compare our derived signatures for magnetic reconnection to the temporal evolution of the current density. The number of QXLs is shown in \autoref{fig:time-plot} (magenta curve).

The time evolution of the number of QXLs implies that practically no reconnection is occurring at the beginning of the simulation.
This is because almost no current sheets thin enough to develop X-lines have yet formed. T
hey only start to appear later as current sheets start to thin down.
The small number of QXLs at those early stages is purely due to spurious reconnection events arising from the numerical PIC noise.
As a result, there is a non-negligible median reconnection rate at those early, and high fluctuation levels can be attributed to the small number of those numerically-generated QXLs with reconnection rates that may not be physical. To suppress this noise, the time evolution of the median reconnection rate is therefore smoothed using an exponential moving average with $\alpha=0.1$.

Note that the peak values of the three displayed curves occur at different times.
While the mean reconnection rates reach peak values around 50$\Omega_{ci}^{-1}$ to 60$\Omega_{ci}^{-1}$,
the number of QXLs peaks at much later times, between $t=120\Omega_{ci}^{-1}$ and $t=160\Omega_{ci}^{-1}$.
Meanwhile, the peak values of the RMS current density is reached in between those peaks, as previously mentioned, around $100\Omega_{ci}^{-1}$.

The reasons for this relative lag between peaks for those different signatures of reconnection will be further analyzed in future publications, since it is outside the scope of this work.
Nevertheless, one possible hypothetical scenario is the following:
as the initial current sheets caused by the initial setup (large-scale perturbations) thin down, the median reconnection rate will obviously increase, leading to the first one of the observed peaks near 50$\Omega_{ci}^{-1}$ to 60$\Omega_{ci}^{-1}$, while the number of reconnection events is still relatively low.
Later on, as previously mentioned, we observed the break up of the largest current sheets with high reconnection rates into smaller ones, with secondary X-lines.
They usually have lower reconnection rates as normally observed in reconnection simulations.
As a result, the number of QXLs will keep increasing but the median reconnection rate will go down as the time goes by.
The maximum of the number of QXLs is reached much later, being the last one of the observed peaks ($t=120\Omega_{ci}^{-1}$ to $t=160\Omega_{ci}^{-1}$).
At those times there are many reconnection events with a majority of low reconnection rates compared to earlier times.
It is therefore plausible that the maximum reconnection activity will occur in between the peak of median reconnection rates and the peak number of QXLs, at a time when the number of QXLs go up and the mean reconnection rate goes down.
Indeed, this time could be the peak of the RMS values of the current density ($t=100\Omega_{ci}^{-1}$), which it has normally been associated to the maximum reconnection activity in turbulence simulations.
Note also the strong correlation between the averaged median reconnection rate and the current density.
With a correlation coefficient of $\rho_{XY}=\text{Cov}(X,Y)/\sigma_X\sigma_Y\approx0.98$, the reconnection rate  closely follows the strength of the current density, but with a clear time delay.
They start to diverge when the number of secondary X-lines with lower reconnection rates becomes dominant.

\subsection{Estimation of the reconnection rate in turbulent plasma}
\label{sec:estimation-reconnection-rate}
\begin{figure}
	\centering
	\includegraphics[width=0.99\linewidth]{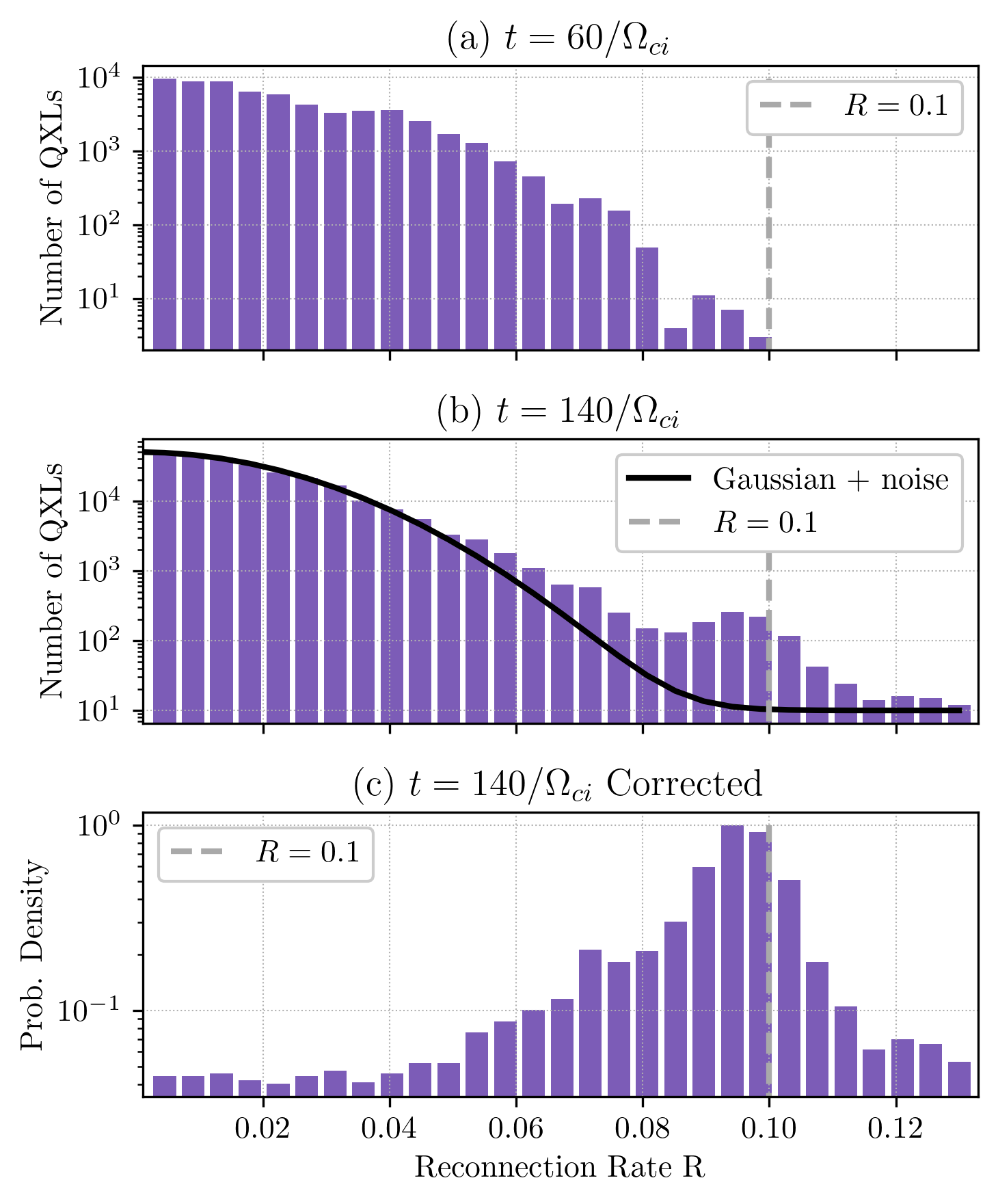}
	\caption{Histogram of the normalized reconnection rate distribution over all extracted quasi X-lines.
		(Left) Distribution close to beginning of reconnection (and peak of mean reconnection rates) at time step $t=60\Omega_{ci}^{-1}$.
		(Right) Distribution at peak of number of X-lines at time step $t=140\Omega_{ci}^{-1}$.}
	\label{fig:reconnection-rate}
\end{figure}

As discussed in Section~\ref{sec:relative-reconnection-rate}, a key advantage of our approach is that it yields a direct estimate of the reconnection rate.
To demonstrate its applicability in a turbulent regime, we analyze the hybrid-kinetic solar wind simulation introduced in Section~\ref{sec:solar-wind}.
\autoref{fig:reconnection-rate} shows the histogram of normalized reconnection rates $R$ measured along all quasi-X-lines for (a) time $t=60\Omega_{ci}^{-1}$, near the first peak of reconnection signatures (mean reconnection rates) and (b) at $t=140\Omega_{ci}^{-1}$, near the third peak of reconnection signatures (number of QXLs).
These times have been chosen according to the time evolution discussed in the previous Section~\ref{sec:temporal-onset-of-magnetic-reconnection}.
Each bin indicates the number of X-lines whose estimated rate falls within that interval.
Note that the number of those X-lines does not necessarily represent the number of reconnection events and many of the QXLs can belong to the same reconnection structure.
Furthermore, many of QXLs at the smallest scales can be due to PIC numerical noise or spurious reconnection events.

In direct comparison of the two histograms, the most notable feature is a clear increase in counts (local maximum) around the 0.1 value for $t=140\Omega_{ci}^{-1}$, i.e., where the number of reconnection events peaks.
At earlier times, e.g., when the mean reconnection rates peaks and there are much less number of reconnection events (at $t=60\Omega_{ci}^{-1}$), the distribution does not contain such an increase in counts, but the background thresholds at 0.1 and virtually no line has any value above 0.1.

Rates substantially below this 0.1 value are likely due either to genuinely slow reconnection events or, more commonly, to background thermal fluctuations and numerical PIC noise, which can lead to spurious, small-scale magnetic reconnection events, but also numerical uncertainty in the estimation of the reconnection rate.
This noise can overshadow the signal of genuine reconnection and makes further analysis difficult.
To address this issue, we assume a thermal background distribution of reconnection rates $p(R)\propto \exp(-R^2)$ and a constant noise floor $\eta(R)=\eta=\text{const}$.
We fit a corresponding fit function to the lower part histogram, i.e., we fit $a,b\in\mathbb{R}$ with
\begin{equation}
	p(R;a,b)=a\exp\left(-bR^2\right)+\eta
\end{equation}
to the measured data distribution $p_X(R)$ and finally divide
\begin{equation}
	\tilde{p}(R)=p_X(R)/p(R;a,b)
\end{equation}
to get a corrected distribution of reconnection rates.
This procedure dampens the contribution from statistical fluctuations and the corrected distribution hence indicates deviations from thermal background.
The resulting corrected histogram is shown in \autoref{fig:reconnection-rate}~(c), where the peak at 0.1 is now clearly visible, consistent with the behavior reported in earlier studies.

Note that we have not explained why the peak at 0.1 is more visible at this time $t=140\Omega_{ci}^{-1}$.
This is just the time at which the number of reconnection events peaks, but not necessarily the time of maximum reconnection activity in the simulation, which presumably takes places earlier (Section~\ref{sec:temporal-onset-of-magnetic-reconnection}).
We defer an explanation to future work, considering that the very reason why reconnection rates are usually found around 0.1 in simulations and nature is still not fully understood, in general.

\section{Discussion and conclusion}
\label{chap:discussion}
In this work, we presented a numerical and analytical framework to identify 3D magnetic reconnection in plasma simulations by applying techniques from the field of fluid visualization.
Using only the magnetic field data, we showed that reconnection X-lines can be found as bifurcation lines using the parallel vectors operator.
We further introduced quasi X-lines, as a novel approach to find reconnection structures, working very well, even in a turbulent plasma characterized by reconnection sites with significant magnetic guide fields.
Additionally, we introduced the $I_2$ shear layers as a complementary tool to diagnose the magnetic field of complex plasma environments.

We successfully applied our framework to three relevant astrophysical plasma scenarios: 3D plasma simulations of a Harris current sheet, a solar flare reconnection and of turbulence-driven reconnection, dominated by guide field reconnection.
These three scenarios were modeled with different plasma models to further illustrate the broad scope of our method: fully kinetic, resistive MHD and hybrid-kinetic, respectively.
Our identification method not only agrees with the results of previously published established methods, but it also revealed new features.

In the Harris sheet simulation, our method could locally extract the X-type singular field lines, as well as quasi-separatrix layers as bifurcation lines. For the coronal flux rope scenario, our method could identify both, the X-type singular field lines and additionally the O-type vortex core lines corresponding to flux ropes.
For the turbulence-driven reconnection case, our method could further identify the time dependence of reconnection events, i.e., the time at which reconnection events start to occur in turbulence and when they reach the maximum activity, etc.
Our approach is independent on the physical scale, since it just relies on the local values of the magnetic field.
So it does not only apply to standard reconnection events with ion coupling, but also to the recently discovered electron-only reconnection events which can develop in turbulence \citep{Phan2018}.

Since the purpose of this paper was the presentation and verification of our methods, in future work we want to focus more on the physics of magnetic reconnection using the tools we developed in this work.
For example, our algorithm can be efficiently used to acquire statistics of reconnection in turbulence including time-dependence, in contrast to previous methods.
In particular, we want to gather statistics of quantities such as the reconnection rate, the properties of the diffusion region like the electron vs\@. ion heating, the energy transfer between field and plasma/dissipation, the number and distribution of plasmoids (O-points identified via vortex core lines), standard versus electron-only reconnection events, etc.
Coupled with our current sheet identification method using the $I_2$ shear layers, our method can also be used to determine statistics of reconnecting vs.
non-reconnecting current sheets in turbulence.

A critical issue across the algorithms is their vulnerability to noise, requiring data smoothing that may obscure significant small-scale details.
Noise is an intrinsic feature, in particular for Lagrangian or semi-Lagrangian methods like kinetic PIC simulations.
The smoothing compromises the balance between computational efficiency and accuracy, especially in high-resolution data scenarios.
Without smoothing, finite-difference approximations of derivatives, integrals, and numerical linear algebra algorithms, can lead to accumulation of errors in the calculations of the location of X-lines, but also in the estimation of derived quantities like the reconnection rate.

Despite the known caveats, our proposed method demonstrates robust and efficient performance across a variety of three-dimensional plasma simulation types and astrophysical configurations.
Notably, it enables the systematic extraction and quantification of magnetic reconnection processes even in complex, time-dependent turbulent regimes, offering a significant advantage over previous approaches that were limited in scope or dimensionality.

All this allows to determine the influence of reconnection on plasma turbulence
and its consequent contribution to the dissipation at the end of the energy turbulent cascade,
one of the most important unsolved questions in space plasma physics.

\section*{SUPPLEMENTARY MATERIAL}
See the supplementary material for a video of X-lines in the solar wind simulation.

\begin{acknowledgments}
	The research of FS has been funded by the Deutsche Forschungsgemeinschaft (DFG, German Research Foundation) SP 1124/9.

	Some computations were performed at the Max Planck Computing and Data Facility (MPCDF).

	The authors gratefully acknowledge the data storage service SDS@hd supported by the Ministry of Science, Research and the Arts Baden-Württemberg (MWK) and the German Research Foundation (DFG) through grant INST 35/1503-1 FUGG.

	The authors acknowledge support by the High Performance and Cloud Computing Group at the Zentrum für Datenverarbeitung of the University of Tübingen, the state of Baden-Württemberg through bwHPC and the German Research Foundation (DFG) through grant no INST 37/935-1 FUGG.

	We would like to extend our sincere gratitude to Yulei Wang \textit{et al.} for providing the simulation data that were essential in comparing and validating our methods.
\end{acknowledgments}

\section*{Author Declarations}

\subsection*{Conflict of Interest}

The authors have no conflicts to disclose.

\subsection*{Author Contributions}

\textbf{Maximilian M. Richter:}
Conceptualization (equal);
Data curation (lead);
Formal analysis (lead);
Investigation (equal);
Methodology (lead);
Resources (lead);
Software (lead);
Validation (lead);
Visualization (lead);
Writing - original draft (equal);
Writing - review \& editing (equal).

\textbf{Patricio A. Mu\~noz:}
Conceptualization (equal);
Investigation (equal);
Software (supporting);
Writing - original draft (equal);
Writing - review \& editing (equal).

\textbf{Felix Spanier:}
Funding acquisition (lead);
Project administration (lead);
Supervision (lead);
Writing - original draft (equal);
Writing - review \& editing (equal).

\section*{Data Availability Statement}

The data that support the findings of this study are available from the corresponding author upon reasonable request.
Source code of the algorithms are publicly available and can be accessed here: \url{https://github.com/maxawake/mrvis}

\appendix
\section{Validation of quasi X-lines}
\label{sec:validation}
\begin{figure}
	\centering
	\begin{tabular}{ccc}
		$B_0=0.1$                                                     & $B_0=0.5$ & $B_0=1.0$ \\
		\includegraphics[width=0.3\linewidth]{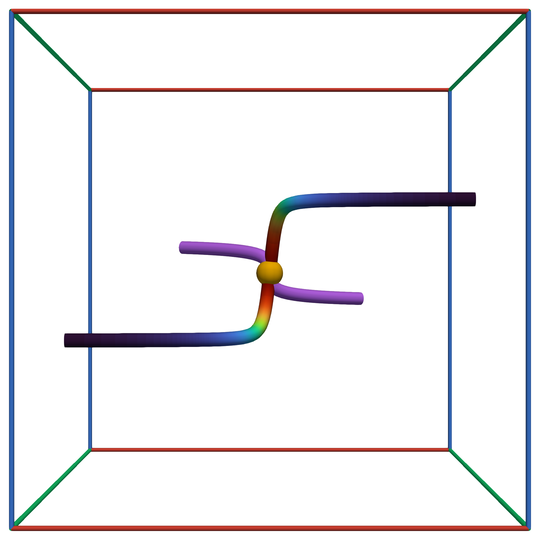} &
		\includegraphics[width=0.3\linewidth]{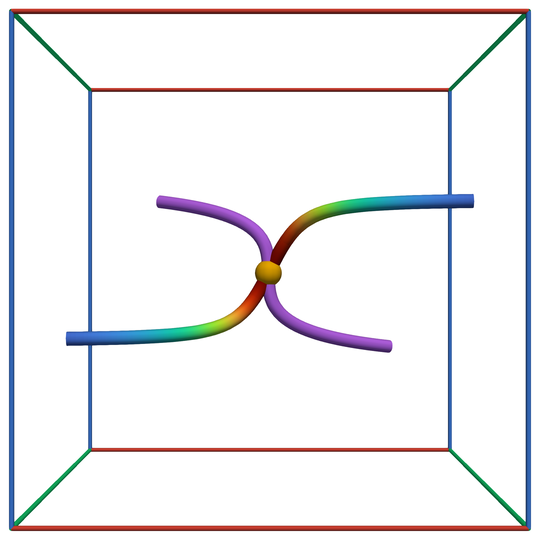} &
		\includegraphics[width=0.3\linewidth]{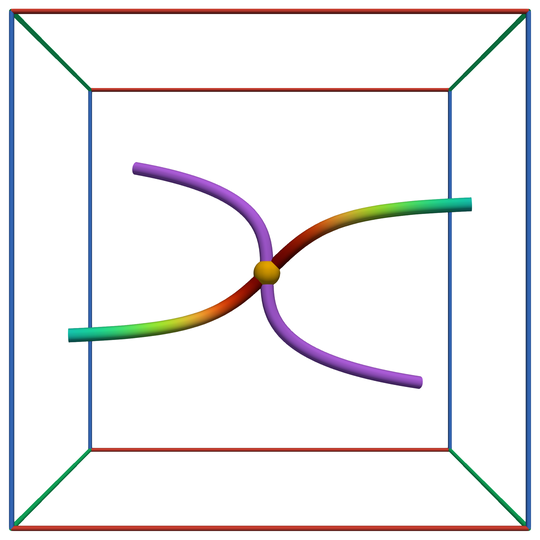}
	\end{tabular}
	\caption{Quasi X-line of the solar flux rope model (\autoref{eq:solar-flux-rope}) for different guide field strengths $B_0$.
		Unfiltered parallel vector lines are shown in turbo rainbow color scheme, showing the feature strength along the line.
		Blue is low feature strength, red is high feature strength.
		Quasi X-lines (purple) are seeded as field lines from the point of the highest feature strength (yellow sphere).
		No further filtering has been applied.}
	\label{fig:qxl}
\end{figure}

The extraction of quasi X-lines from Section~\ref{sec:qxls} can be justified with the twisted solar flux rope model (\autoref{eq:solar-flux-rope}).
The unfiltered solutions of the parallel vectors operator for the flux rope model for different guide field strengths are shown in \autoref{fig:qxl}, where the feature strength $\phi$ of the solution line is colored in the turbo rainbow color scheme.
The QXL (shown in purple) is extracted straightforwardly as a field line of the vector field, integrated from the point of the highest hyperbolicity (shown as a yellow sphere) along the parallel vector line.
Increasing the guide field $B_0$, we find that the hyperbolic flow is not the dominant behavior along the bifurcation line anymore.
The hyperbolic flow and the line structure still exist, but the line is not a solution of the parallel vectors operator anymore.
In such cases, QXLs should be used.

\begin{figure}
	\centering
	\includegraphics[width=0.45\linewidth]{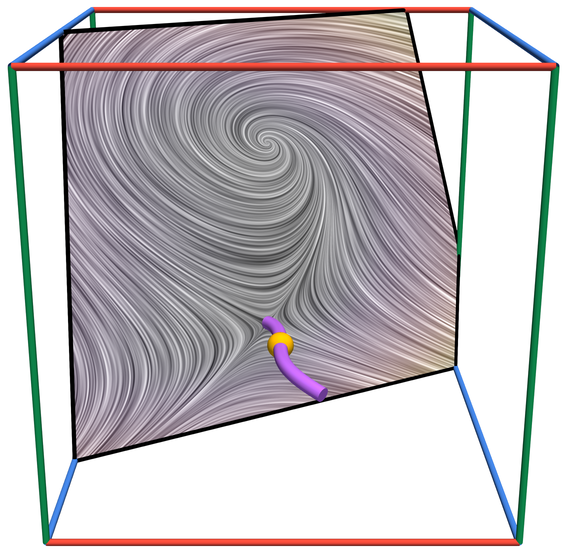}
	\includegraphics[width=0.45\linewidth]{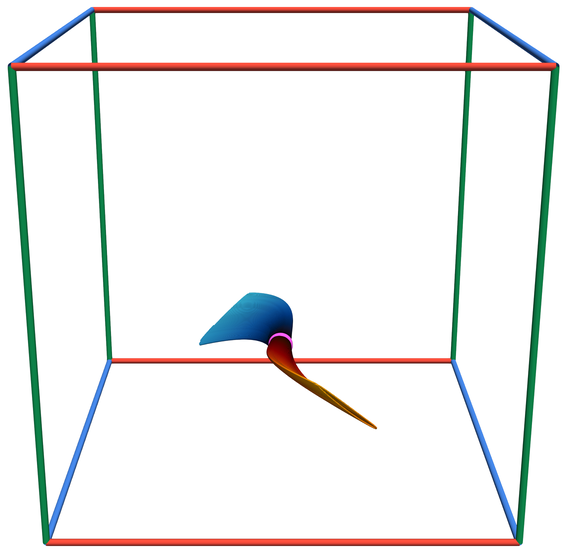}
	\caption{(Left) Slice of the solar flux rope model perpendicular to the quasi X-line (purple).
		The projected vector field onto the slice plane is visualized as a LIC.
		(Right) Stream tube around the quasi X-line.
		The stream tube shows the two stable manifolds of forward (red) and backward (blue) integration in time.
		The seed points of the stream tube are shown in pink.}
	\label{fig:qxl-filtered}
\end{figure}

If we further filter the QXL with the bifurcation line criterion (\autoref{eq:feature-strength}), we find field lines of the vector field showing the desired hyperbolic properties of X-lines, even in cases of large guide fields $B_0$.
This is illustrated in \autoref{fig:qxl-filtered}, where the quasi X-line (purple) shows X-points in the in-plane vector field.
This is further emphasized in \autoref{fig:qxl-filtered} by a small tube of field lines around the QXL, showing the two stable and flat manifolds of forward and backward integration in time.

The definition of QXLs is not restricted to hyperbolic manifolds and can also be expanded for vortex core lines.
It can be applied to any vector field, where the hyperbolic behavior is not the dominant component of the field.

\bibliography{aipsamp}
\end{document}